\documentclass[pre,twocolumn,superscriptaddress,showpacs]{revtex4-1}
\usepackage{graphicx,epsfig,subfigure}
\usepackage{amsbsy}
\usepackage{amssymb}
\usepackage{amsmath}
\usepackage{bm}
\usepackage{lmodern}
\usepackage[T1]{fontenc}
\usepackage[utf8]{inputenc}
\usepackage{hyperref}
\usepackage[bottom]{footmisc}
\usepackage{xcolor}
\usepackage{ulem}
\newcommand{\eqnref}[1]{Eq.~(\ref{#1})}
\newcommand{\figref}[2]{Fig.~\ref{#2}#1}
\newcommand{\beq}{\begin{equation}}
\newcommand{\eeq}{\end{equation}}
\newcommand{\beqa}{\begin{eqnarray}}
\newcommand{\eeqa}{\end{eqnarray}}

\newcommand{\ds}[2]{\frac{\dd #1}{\dd #2}}
\newcommand{\pd}[2]{\frac{\partial #1}{\partial #2}}
\newcommand{\half}{\frac{1}{2}}
\newcommand{\rk}[1]{\left( #1 \right)}
\newcommand{\sk}[1]{\left[ #1 \right]}
\newcommand{\bk}[1]{\left\lbrace #1 \right\rbrace}

\newcommand{\ii}{\mathrm{i}}
\newcommand{\ee}{\mathrm{e}}
\newcommand{\dd}{\mathrm{d}}

\newcommand{\bz}{k_{\text{B}}}

\date{\today}

\begin{document}

\title{Quantifying the validity and breakdown of the overdamped approximation in stochastic thermodynamics: Theory and experiment}

\author{Rui Pan}
\affiliation{School of Physics, Peking University, Beijing 100871, People's Republic of China}

\author{Thai M. Hoang}
\thanks{Current address: Sandia National Laboratories, Albuquerque, New Mexico 87123, USA}
\affiliation{Department of Physics and Astronomy, Purdue University, West Lafayette, Indiana 47907, USA} 

\author{Zhaoyu Fei}
\affiliation{School of Physics, Peking University, Beijing 100871, People's Republic of China}

\author{Tian Qiu}
\affiliation{School of Physics, Peking University, Beijing 100871, People's Republic of China}

\author{Jonghoon Ahn}
\affiliation{School of Electrical and Computer Engineering, Purdue University, West Lafayette, Indiana 47907, USA}

\author{Tongcang Li}
\affiliation{Department of Physics and Astronomy, Purdue University, West Lafayette, Indiana 47907, USA}
\affiliation{School of Electrical and Computer Engineering, Purdue University, West Lafayette, Indiana 47907, USA}
\affiliation{Purdue Quantum Center, Purdue University, West Lafayette, Indiana 47907, USA}
\affiliation{Birck Nanotechnology Center, Purdue University, West Lafayette, Indiana 47907, USA}

\author{H. T. Quan}
\email{Corresponding author: htquan@pku.edu.cn}
\affiliation{School of Physics, Peking University, Beijing 100871, People's Republic of China}
\affiliation{Collaborative Innovation Center of Quantum Matter, Beijing 100871, People's Republic of China}

\begin{abstract}
{
Stochastic thermodynamics provides an important framework to explore small physical systems where thermal fluctuations are inevitable. 
In the studies of stochastic thermodynamics, some thermodynamic quantities, such as the trajectory work, associated with the complete Langevin equation (the Kramers equation) are often assumed to converge to those associated with the overdamped Langevin equation (the Smoluchowski equation) in the overdamped limit under the overdamped approximation. 
Nevertheless, a rigorous mathematical proof of the convergence of the work distributions to our knowledge has not been reported so far. Here we study the convergence of the work distributions explicitly. 
In the overdamped limit, we rigorously prove the convergence of the extended Fokker-Planck equations including work using a multiple timescale expansion approach. By taking the linearly dragged harmonic oscillator as an exactly solvable example, we analytically calculate the work distribution associated with the Kramers equation, and verify its convergence to that associated with the Smoluchowski equation in the overdamped limit. We quantify the accuracy of the overdamped approximation as a function of the damping coefficient. In addition, 
we experimentally demonstrate that the data of the work distribution of a levitated silica nanosphere agrees with the overdamped approximation in the overdamped limit, but deviates from the overdamped approximation in the low-damping case. Our work 
fills a gap between the stochastic thermodynamics based on the complete Langevin equation (the Kramers equation) and the overdamped Langevin equation (the Smoluchowski equation), and deepens our understanding of the overdamped approximation 
in stochastic thermodynamics.
}
\end{abstract}

\maketitle

\section{Introduction}


The Brownian motion was named after Robert Brown when he observed jiggling motions of tiny particles suspended in water under a microscope in 1827. Brown immediately realized that the motion was not life-related. However, it was not until the beginning of the 1900s when Einstein, Smoluchowski, and Langevin developed the breakthrough theories that precisely explained the origin of the Brownian motion. In general, particles exhibit the Brownian motion due to their collisions with other particles, where the random forces acting on the particles result from thermal fluctuations of the environment. While the Langevin equation (LE) is often used to describe the temporal dynamics of a Brownian system (e.g., the colloidal and biomolecular systems \cite{purcell1977life}), the LE description is equivalent to the Fokker-Planck equation description \cite{risken1996fokker}. The complete Fokker-Planck equation, corresponding to the complete LE, is known as the Kramers equation (KE) \cite{kramers1940brownian}. In the overdamped limit, the LE is simplified to a first-order stochastic differential equation using the overdamped approximation (OA), where the particle inertia is negligible compared to the damping force. The overdamped Fokker-Planck equation, corresponding to the overdamped LE, is the Smoluchowski equation (SE) \cite{Smoluchowski1916brownian}. A rigorous mathematical proof of the convergence of the KE to the SE in the overdamped limit was firstly given in the 1970s using systematic expansion procedures \cite{wilemski1976derivation, titulaer1978systematic, van1992stochastic, bocquet1997high}.

The past two decades have witnessed significant developments in nonequilibrium statistical mechanics of small systems where thermal fluctuations dominate \cite{jarzynski1997nonequilibrium, crooks1999entropy, sekimoto1998langevin, sekimoto2010stochastic, seifert2005entropy, seifert2012stochastic, liphardt2002equilibrium, collin2005verification, douarche2006work, junier2009recovery, martinez2016brownian, toyabe2010experimental}. At the microscale level, thermodynamic quantities such as work, heat \cite{sekimoto1998langevin, sekimoto2010stochastic}, and entropy production \cite{seifert2005entropy} have been defined and studied along individual stochastic trajectories. Based on these work, fluctuation theorems were derived \cite{jarzynski1997nonequilibrium, crooks1999entropy, seifert2005entropy, seifert2012stochastic}, giving rise to the emerging field of stochastic thermodynamics. In previous studies of stochastic thermodynamics, attentions are predominantly focused on the high-damping regime, and people have been using the OA in calculating the distribution of thermodynamic quantities. 
Most recently, the inertial effects and asymptotic behaviors in the overdamped limit have attracted a lot of attentions \cite{celani2012anomalous, bo2013entropic, bo2016multiple, ge2014time, martinez2015adiabatic, kawaguchi2013fluctuation, lan2015stochastic, marino2016entropy, taniguchi2008inertial, kwon2013work, imparato2006fluctuation, gomez2008optimal, baule2009steady, minh2009path, engel2009asymptotics, holubec2015asymptotics, li2013brownian, esposito2015stochastic, spinney2012entropy, kim2004entropy, zhao2017brownian, esposito2012stochastic, li2017shortcuts, yin2006existence}, and researchers begin to examine the validity of the OA in stochastic thermodynamics. 
For example, the behaviors of entropy production have been studied explicitly \cite{celani2012anomalous, bo2013entropic, bo2016multiple, martinez2015adiabatic, kawaguchi2013fluctuation, lan2015stochastic, marino2016entropy}. Especially, it is found that in the presence of a spatial \cite{celani2012anomalous, bo2016multiple} or temporal \cite{martinez2015adiabatic} dependent temperature profile, the OA fails to capture the correct behaviors of entropy production. In Refs.~\cite{bo2013entropic, ge2014time, bo2016multiple}, it has been shown that the heat distribution associated with the KE converges to that associated with the SE in the overdamped limit. But the validity of the OA for the work distribution to our knowledge has not been directly justified so far. Also, the breakthrough in experimental techniques of measuring the instantaneous velocity of the Brownian particle \cite{li2010measurement, kheifets2014observation} enables us to study the validity of the OA in stochastic thermodynamics systematically.

In Ref.~\cite{bocquet1997high}, by utilizing a multiple timescale approach \cite{bocquet1997high, wang2016entropy, celani2012anomalous, bo2016multiple}, it has been demonstrated that the marginal distribution of the solution to the KE in the position space converges to the solution to the SE in the overdamped limit. In the current paper, we investigate how the OA affects the thermodynamics besides the dynamics. We study the joint probability distribution in the position and \emph{work} space associated with the KE using the same method, and we find that, in the overdamped limit, the joint distribution converges to 
that associated with the SE. 
Then, as an exactly solvable example, we analytically calculate the joint probability distribution of the position, the velocity, and work for the linearly dragged harmonic oscillator with inertia, and verify its convergence to the solution associated with the corresponding SE \cite{mazonka1999exactly, imparato2007work, taniguchi2007onsager, speck2005dissipated} in the overdamped limit. We use the relative error between the exact solution and the OA to quantify the accuracy of the OA as a function of the damping coefficient. It is found that the OA leads to a deviation of 10\% from the exact solution when the damping coefficient is 2.5 times of the trapping frequency, and 3\% for 5 times of the trapping frequency. We also provide the experimental results which always agree with the solution to the KE, and deviate from the solution to the SE in the low-damping case. These results deepen our understanding of the OA 
in stochastic thermodynamics.

This paper is organized as follows. In Sec. \ref{Sec:MTSA}, as a general case, we give the rigorous mathematical proof of the convergence of the work distributions in the overdamped limit. In Sec. \ref{Sec:LDHO}, as an exactly solvable example, we analytically calculate the work distribution of the linearly dragged harmonic oscillator with inertia, and verify its convergence to the work distribution associated with the SE quantitatively. In Sec. \ref{Sec:Experiment}, we provide the experimental results which verify the predictions in Sec. \ref{Sec:LDHO}. Section \ref{Sec:Conclusion} presents our conclusions.

\section{\label{Sec:MTSA}The convergence of work: A multiple timescale approach}

For simplicity, in this paper, we consider the case in which both the temperature $T$ and the friction coefficient $\gamma$ are homogeneous in space and time. The motion [trajectory in the phase space ($x$,$v$)] of the Brownian particle is described by the complete LE:
\beqa
\begin{split}
\dd x &= v \dd t,\\
\dd v &= \left( -\Gamma v + \frac{\mathcal{F}(x,t)}{m} \right) \dd t + \sqrt{\frac{2\Gamma}{\beta m}}\dd \mathcal{B}_t,
\end{split}
\label{Eqn:ULE0}
\eeqa
where $m$ is the mass of the particle, $\Gamma := \gamma/m$ is the damping coefficient, $\mathcal{F}(x,t) = -\partial{\mathcal{U}(x,t)}/\partial{x} + \mathcal{F}_{nc}(x,t)$ is the external forces including the contribution from the potential $\mathcal{U}(x,t)$ and the nonconservative force $\mathcal{F}_{nc}(x,t)$, $\beta = \rk{\bz T}^{-1}$ is the inverse temperature, $\bz$ is Boltzmann's constant, and $\mathcal{B}_t$ describes the standard Brownian motion (Wiener process) \cite{sekimoto2010stochastic}.

In stochastic thermodyamics, work is defined as a functional along individual trajectories \cite{sekimoto1998langevin, seifert2012stochastic, celani2012anomalous}
\beq
\dj w\sk{x(t)} = \pd{\mathcal{U}}{t}\dd t + \mathcal{F}_{nc} \circ \dd x,
\label{WorkDefinition}
\eeq
where $\circ$ denotes the Stratonovich product. The work functional corresponding to the complete LE [\eqnref{Eqn:ULE0}] can be expressed as
\beq
\dj w = \left( \pd{\mathcal{U}}{t} + \mathcal{F}_{nc} v \right) \dd t.
\label{Eqn:ULEW0}
\eeq
From the point of view of the probability distribution, the extended Kramers equation (EKE) including work reads \cite{risken1996fokker, imparato2005work}
\begin{widetext}
\beq
\pd{}{v} \left( v+ \frac{1}{\beta m} \pd{}{v}\right) p(x,v,w;t) =
\Gamma^{-1} \left[ \pd{}{t} + v\pd{}{x}+ \frac{\mathcal{F}}{m} \pd{}{v} + \left( \pd{\mathcal{U}}{t} + \mathcal{F}_{nc} v \right) \pd{}{w} \right] p(x,v,w;t),
\label{Eqn:UFPW0}
\eeq
\end{widetext}
where $p(x,v,w;t)$ is the joint probability distribution of the position, the velocity, and work at time $t$ associated with the EKE [\eqnref{Eqn:UFPW0}].

In the overdamped limit ($\Gamma \rightarrow \infty$), the complete LE is reduced to the overdamped LE by taking the OA ($m \ddot x \rightarrow 0$):
\beq
\dd x = \frac{1}{\gamma}\mathcal{F}(x,t)\dd t + \sqrt{\frac{2}{\beta \gamma}}\dd \mathcal{B}_t.
\label{Eqn:OLE0}
\eeq
According to \eqnref{WorkDefinition}, the corresponding work functional can be written as
\beqa
\dj w = & & \left[ \pd{\mathcal{U}}{t} + \Gamma^{-1} \frac{1}{m} \left( \mathcal{F} \mathcal{F}_{nc} + \frac{1}{\beta} \mathcal{F}_{nc}' \right) \right] \dd t \nonumber\\
& & + \sqrt{\frac{2 \Gamma^{-1}}{\beta m}} \mathcal{F}_{nc} \cdot \dd \mathcal{B}_t,
\label{Eqn:OLEW0}
\eeqa
where the prime denotes a derivative with respect to $x$, and $\cdot$ denotes the It\^o product. The extended Smoluchowski equation (ESE) including work reads \cite{risken1996fokker, imparato2005work}
\begin{widetext}
\beq
\pd{}{t}\rho(x,w;t) = \Gamma^{-1} \frac{1}{m} \rk{\pd{}{x} + \mathcal{F}_{nc} \pd{}{w}} \rk{\frac{1}{\beta} \pd{}{x} - \mathcal{F} + \frac{1}{\beta} \mathcal{F}_{nc} \pd{}{w}} \rho(x,w;t) - \pd{\mathcal{U}}{t} \pd{\rho}{w},
\label{Eqn:OFPW0}
\eeq
\end{widetext}
where $\rho(x,w;t)$ is the joint probability distribution of the position and work at time $t$ associated with the ESE [\eqnref{Eqn:OFPW0}]. The derivations of Eqs.~(\ref{Eqn:OLEW0}) and (\ref{Eqn:OFPW0}) are given in Appendix \ref{A:OverdampedEqns}.

In the overdamped limit ($\Gamma^{-1} \rightarrow 0$), since the Brownian particle relaxes in very short timescale ($\tau_p = \Gamma^{-1}$) in the velocity space in comparison with that in the position space, we can separate timescales of the evolutions in the position and the velocity spaces. This is the so-called multiple time-scale approach \cite{bocquet1997high, wang2016entropy, celani2012anomalous, bo2016multiple}. By utilizing this approach, we rigorously prove the convergence of \eqnref{Eqn:UFPW0} to \eqnref{Eqn:OFPW0} in the overdamped limit (see Appendix \ref{A:MTSA}). In other words, if we define the marginal distribution of $p(x,v,w;t)$ as
\beq
\tilde{p}(x,w;t) = \int p(x,v,w;t) \dd v,
\label{Sol:XWpdf0}
\eeq
we find that in the overdamped limit, the equation of motion for $\tilde{p}(x,w;t)$ can be written as
\begin{widetext}
\beqa
\pd{}{t}\tilde{p}(x,w;t) = \Gamma^{-1} \frac{1}{m} \rk{\pd{}{x} + \mathcal{F}_{nc} \pd{}{w}} \rk{\frac{1}{\beta} \pd{}{x} - \mathcal{F} + \frac{1}{\beta} \mathcal{F}_{nc} \pd{}{w}} \tilde{p}(x,w;t) - \pd{\mathcal{U}}{t} \pd{\tilde{p}}{w} + \mathcal{O}\rk{\Gamma^{-2}}.
\label{Eqn:OFPW02}
\eeqa
\end{widetext}
It can be seen that \eqnref{Eqn:OFPW02} is exactly the same as \eqnref{Eqn:OFPW0} when the higher-order corrections are ignored in the overdamped limit.

We conclude that \eqnref{Eqn:UFPW0} converges to \eqnref{Eqn:OFPW0} in the limit $\Gamma^{-1} \rightarrow 0$, that is to say \cite{arrow},
\beq
\int p(x,v,w;t) \dd v \ \rightarrow \ \rho(x,w;t).
\label{Eqn:Convergence}
\eeq
More precisely,
\beq
\int p(x,v,w;t) \dd v = \rho(x,w;t) + \mathcal{O}\rk{\Gamma^{-2}}.
\label{Eqn:Expansion}
\eeq
As a result, both marginal distributions of the position and work from solving the EKE converge to those from solving the ESE respectively as long as the initial conditions are the same. Here we would like to emphasize that 
in our proof, we have assumed the temperature and the friction to be constants in space and time, but our results are still valid even when there is a spatial or temporal dependent friction and temperature. Hence the validity of the OA in calculating the work distribution in stochastic thermodynamics is analytically confirmed in the overdamped limit.

Before concluding this section, we would like to give the following remark. 
In previous work, the thermodynamic quantities of heat \cite{bo2013entropic, ge2014time}, entropy production \cite{celani2012anomalous, kawaguchi2013fluctuation, lan2015stochastic, marino2016entropy} and other functionals \cite{bo2016multiple} were investigated using the same method. The convergence of the overall heat released to the environment was confirmed in the overdamped limit \cite{bo2013entropic, ge2014time}. By combining the first law of stochastic thermodynamics and the convergence of the distributions of heat and the internal energy, the convergence of work distributions can be anticipated while here we explicitly show the expected result in a direct way.

In the following section, we will quantify the accuracy of the OA as a function of the damping coefficient in a harmonic system.

\section{\label{Sec:LDHO}An exactly solvable model: Linearly dragged harmonic oscillator}

\subsection{The model and the extended Kramers equation}

In the preceding section, we give the rigorous proof of the convergence of work distributions for a generic system without solving the EKE and the ESE. In this section, as an example, we consider the model of a dragged harmonic oscillator with a constant dragging speed. This model is exactly solvable \cite{mazonka1999exactly, imparato2007work, taniguchi2007onsager, speck2005dissipated}. A particle is doing Brownian motion in a heat bath whose inverse temperature is $\beta = \rk{\bz T}^{-1}$. Meanwhile, the Brownian particle is subject to the driving force of a harmonic trap whose center is shifted linearly in time. 
The potential can be described by
\beq
\mathcal{U}(x,t) = \half m \omega^2 \rk{x-ut}^2,
\label{HO:Potential}
\eeq
where $\omega$ is the trapping frequency of the harmonic potential, and $u$ is the speed of the shifting of the potential center. According to \eqnref{WorkDefinition}, the time derivative of the work functional can be written as
\beq
\frac{\dj w}{\dd t} = \pd{\mathcal{U}(x,t)}{t} = - m u \omega^2 \rk{x-ut}.
\label{HO:Work}
\eeq
For convenience, we transfer to the reference frame of the shifting potential, and define the relative position with respect to the center of the potential by $y = x-ut$.

We start from the complete LE [\eqnref{Eqn:ULE0}] and its corresponding work functional [\eqnref{WorkDefinition}],
\beqa
\begin{split}
& \dd y = \rk{v - u} \dd t, \\
& \dd v = \rk{- \Gamma v - \omega^2 y} \dd t + \sqrt{\frac{2 \Gamma}{\beta m}} \dd \mathcal{B}_t, \\
& \dj w = - m u \omega^2 y \ \dd t.
\end{split}
\label{HO:ULEW}
\eeqa
The corresponding EKE about the joint probability distribution $p(y,v,w,t)$ of the position $y$, the velocity $v$, and work $w$ at time $t$ can be written as
\beqa
\pd{p}{t} = & & \rk{u - v} \pd{p}{y} + \Gamma \rk{p + v\pd{p}{v}} + \omega^2 y \pd{p}{v} \nonumber\\
& & + \frac{\Gamma}{\beta m} \frac{\partial^2 p}{\partial v^2} + m u \omega^2 y \pd{p}{w}.
\label{HO:UFPW}
\eeqa
Let the initial state be a Gaussian distribution. The solution of $p(y,v,w,t)$ remains a Gaussian distribution because the drift and the diffusion coefficients in \eqnref{HO:UFPW} are either constant or linear in $y$, $v$, and $w$ \cite{mazonka1999exactly, kwon2013work}. By utilizing the Gaussian property, $p(y,v,w,t)$ can be written in the form of a three-dimensional Gaussian distribution
\beq
p^G(y,v,w,t) = \frac{1}{(2 \pi)^{\frac{3}{2}} | \bm{\Sigma} |^{\half}} \exp \rk{- \half \bm{z}^T \bm{\Sigma}^{-1} \bm{z}},
\label{HO:Gaussian}
\eeq
where 
\beq
\bm{z} = \left(
\begin{array}{c}
y - \bar{y} \\
v - \bar{v} \\
w - \bar{w}
\end{array} \right),
\ \bm{\Sigma} = \left(
\begin{array}{ccc}
\sigma^2_y & c_{yv} & c_{yw} \\
c_{yv} & \sigma^2_v & c_{vw} \\
c_{yw} & c_{vw} & \sigma^2_w
\end{array} \right),
\label{HO:Matrix}
\eeq
$\bar{y}$, $\bar{v}$, $\bar{w}$ are mean values, $\sigma^2_y$, $\sigma^2_v$, $\sigma^2_w$ are variances, and $c_{yv}$, $c_{yw}$, $c_{vw}$ are the covariances of any two variables of the position, the velocity, and work. These nine moments are functions of time $t$ only, and they determine the evolution of the distribution $p^G(y,v,w,t)$.

Inserting the definitions of the nine moments \cite{definition} to \eqnref{HO:UFPW}, we obtain a set of coupled equations of motion for the nine moments:
\begin{subequations}
\beqa
& & \ds{}{t} \bar{y} = \bar{v} - u,
\label{EOM:a}\\
& & \ds{}{t} \bar{v} = - \Gamma \bar{v} - \omega^2 \bar{y},
\label{EOM:b}\\
& & \ds{}{t} \bar{w} = - m u \omega^2 \bar{y},
\label{EOM:c}\\
& & \ds{}{t} \sigma^2_y = 2 c_{yv},
\label{EOM:d}\\
& & \ds{}{t} \sigma^2_v = - 2 \Gamma \sigma^2_v - 2 \omega^2 c_{yv} + \frac{2\Gamma}{\beta m},
\label{EOM:e}\\
& & \ds{}{t} \sigma^2_w = -2m u \omega^2 c_{yw},
\label{EOM:f}\\
& & \ds{}{t} c_{yv} = \sigma^2_v - \Gamma c_{yv} - \omega^2 \sigma^2_y,
\label{EOM:g}\\
& & \ds{}{t} c_{yw} = c_{vw} - m u \omega^2 \sigma^2_y,
\label{EOM:h}\\
& & \ds{}{t} c_{vw} = - \Gamma c_{vw} - \omega^2 c_{yw} - m u \omega^2 c_{yv}.
\label{EOM:i}
\eeqa
\label{HO:EOM}
\end{subequations}
In our current study, we consider the equilibrium initial state:
\beq
p^G(y,v,w,0|eq) = p^G(y)p^G(v)\delta(w),
\label{InitialState:eq}
\eeq
i.e., at the initial moment of time, $\bar{y}=\bar{v}=\bar{w}=\sigma^2_w=c_{yv}=c_{yw}=c_{vw}=0$, $\sigma^2_y = \frac{1}{\beta m \omega^2}$, $\sigma^2_v = \frac{1}{\beta m}$. In the following, we will solve \eqnref{HO:EOM} to obtain the time evolution of these moments.

\subsection{\label{SubSec:EKE}Solution to the extended Kramers equation [\eqnref{HO:UFPW}]}

From Eqs.~(\ref{EOM:a}), (\ref{EOM:b}), and (\ref{EOM:c}), we have
\beq
\ddot{\bar{y}} + \Gamma \dot{\bar{y}} + \omega^2 \bar{y} = - u \Gamma,
\label{Couple:1}
\eeq
where the dot above the variable denotes the time derivative, and double dots denote the second-order derivative with respect to time. This ordinary differential equation (ODE) has two characteristic roots:
\beq
\mu_{1,2} = -\frac{\Gamma}{2} \pm \sqrt{\rk{\frac{\Gamma}{2}}^2 - \omega^2}.
\label{HO:Mu}
\eeq
It is necessary to clarify some terminologies. Depending on the characteristic roots being complex or real, we define two \emph{regimes}: the low-damping regime (
$\Gamma/\omega < 2$) and the high-damping regime (
$\Gamma/\omega > 2$) \cite{baule2009steady, li2013brownian}. 
The critical damping coefficient between two regimes is $\Gamma_c = 2\omega$. 
In the overdamped limit ($\Gamma/\omega \gg 2$), the characteristic roots $\mu_{1,2}$ converge to
\beqa
\begin{split}
& \mu_1 \rightarrow -\frac{\omega^2}{\Gamma},\\
& \mu_2 \rightarrow - \Gamma,
\end{split}
\label{Limit:Mu}
\eeqa
respectively.

Equations (\ref{Couple:1}) and (\ref{EOM:a}), (\ref{EOM:b}), and (\ref{EOM:c}) give the solutions of the mean values,
\begin{subequations}
\beqa
\bar{y}(t) = &\ & C_1 \ee^{\mu_1 t} + C_2 \ee^{\mu_2 t} - \frac{u\Gamma}{\omega^2},
\label{Solution_1:ybar}\\
\bar{v}(t) = &\ & C_1 \mu_1 \ee^{\mu_1 t} + C_2 \mu_2 \ee^{\mu_2 t} + u,
\label{Solution_1:vbar}\\
\bar{w}(t) = &\ & m u^2 \Gamma t \nonumber\\
&\ & - m u \omega^2 \sk{\frac{C_1}{\mu_1}\rk{\ee^{\mu_1 t}-1} + \frac{C_2}{\mu_2}\rk{\ee^{\mu_2 t}-1}},
\label{Solution_1:wbar}
\eeqa
\label{Solution:1}
\end{subequations}
where the two constants $C_1$ and $C_2$ and their asymptotic values can be determined by the initial condition [\eqnref{InitialState:eq}],
\beqa
C_1 &=& \rk{-u-\frac{u\Gamma}{\omega^2}\mu_2}/\rk{\mu_1 - \mu_2} \rightarrow \frac{u\Gamma}{\omega^2},\nonumber\\
C_2 &=& \rk{u+\frac{u\Gamma}{\omega^2}\mu_1}/\rk{\mu_1 - \mu_2} \rightarrow 0.
\label{C12:Eq}
\eeqa
As a self-consistent check, we can see that the asymptotic values of the mean values [\eqnref{Solution:1}] are
\begin{subequations}
\beqa
& & \bar{y}(t|eq) \rightarrow \frac{u\Gamma}{\omega^2} \rk{\ee^{-\frac{\omega^2}{\Gamma} t}-1},
\label{Solution_Eq_O:ybar}\\
& & \bar{w}(t|eq) \rightarrow \frac{m u^2 \Gamma^2}{\omega^2} \rk{\ee^{-\frac{\omega^2}{\Gamma} t}-1 + \frac{\omega^2}{\Gamma}t},
\label{Solution_Eq_O:wbar}
\eeqa
\label{Solution:1_Eq_O}
\end{subequations}
where the r.h.s. of \eqnref{Solution:1_Eq_O} is identical to the solutions to the ESE obtained in Ref.~\cite{imparato2007work}.

By combining Eqs.~(\ref{EOM:d}), (\ref{EOM:e}), and (\ref{EOM:g}), we obtain a third-order ODE of $\sigma^2_y(t)$:
\beq
\dddot{\sigma^2_y} + 3 \Gamma \ddot{\sigma^2_y} + \rk{4 \omega^2+ 2 \Gamma^2} \dot{\sigma^2_y} + 4 \Gamma \omega^2 \sigma^2_y = \frac{4\Gamma}{\beta m},
\label{Couple:2}
\eeq
where the triple dots denote the third-order time derivative. Three characteristic roots are $\lambda_{1,2,3}$:
\beqa
\begin{split}
\lambda_1 &= 2 \mu_1 \rightarrow -\frac{2 \omega^2}{\Gamma},\\
\lambda_2 &= 2 \mu_2 \rightarrow -2 \Gamma,\\
\lambda_3 &= \mu_1 + \mu_2 =  -\Gamma.
\end{split}
\label{HO:Lambda}
\eeqa

By adapting a similar procedure (see Appendix \ref{A:EqInitial}), we obtain
\begin{subequations}
\beqa
& & \sigma^2_y(t|eq) = \frac{1}{\beta m \omega^2},
\label{Solution_Eq_U:yvar}\\
& & \sigma^2_v(t|eq) = \frac{1}{\beta m},
\label{Solution_Eq_U:vvar}\\
& & c_{yv}(t|eq) = 0,
\label{Solution_Eq_U:cyv}
\eeqa
\label{Solution:2_Eq_U}
\end{subequations}
and
\begin{subequations}
\beqa
& & \sigma^2_w(t|eq) \rightarrow \frac{2m u^2 \Gamma^2}{\beta \omega^2} \rk{\ee^{-\frac{\omega^2}{\Gamma} t}-1 + \frac{\omega^2}{\Gamma}t},
\label{Solution_Eq_O:wvar}\\
& & c_{yw}(t|eq) \rightarrow \frac{u\Gamma}{\beta \omega^2} \rk{\ee^{-\frac{\omega^2}{\Gamma} t}-1}.
\label{Solution_Eq_O:cyw}
\eeqa
\label{Solution:3_Eq_O}
\end{subequations}
We can see that all $\sigma^2_y$, $\sigma^2_v$, and $c_{yv}$ are constants in time. The r.h.s. of Eqs.~(\ref{Solution_Eq_U:yvar}) and (\ref{Solution:3_Eq_O}) is identical to the solutions to the ESE obtained in Ref.~\cite{imparato2007work}.

\begin{figure*}[t]
	\includegraphics[scale=0.42,angle=0]{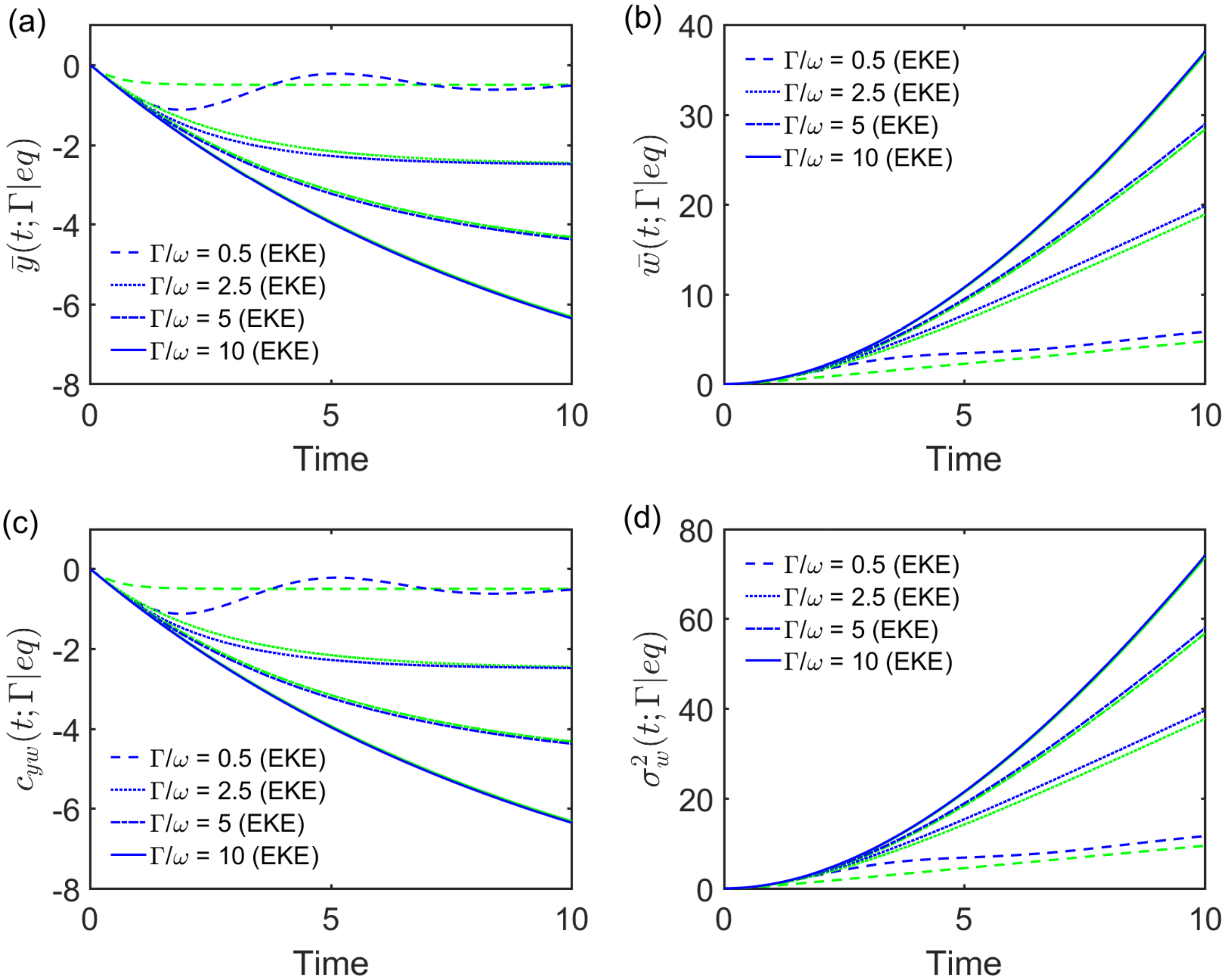}
	\caption{(Color online). Convergence behaviors of various moments [Eqs.~(\ref{Solution:1_Eq_O}) and (\ref{Solution:3_Eq_O})] with different damping coefficients $\Gamma$ for the equilibrium initial state. (a) Mean value of $y$. (b) Mean value of $w$. (c) Covariance of $y$ and $w$. (d) Variance of $w$. Here $\beta = u = m = \omega = 1$. In each figure, there are four pairs of curves corresponding to four damping coefficients ($\Gamma/\omega = 0.5, 2.5, 5, 10$). The critical damping coefficient between the low-damping and the high-damping regimes is $\Gamma_c/\omega = 2$. The solutions to the EKE [blue (dark gray) curves] approach the solutions to the ESE [green (light gray) curves] as the damping coefficient increases. These solutions exhibit an oscillating behavior in the low-damping regime ($\Gamma < \Gamma_c$) while quickly reaching an asymptotic behavior in the high-damping regime ($\Gamma > \Gamma_c$).
	}
	\label{Fig:Analytical_Eq}
\end{figure*}

\begin{figure*}[t]
	\includegraphics[scale=0.48,angle=0]{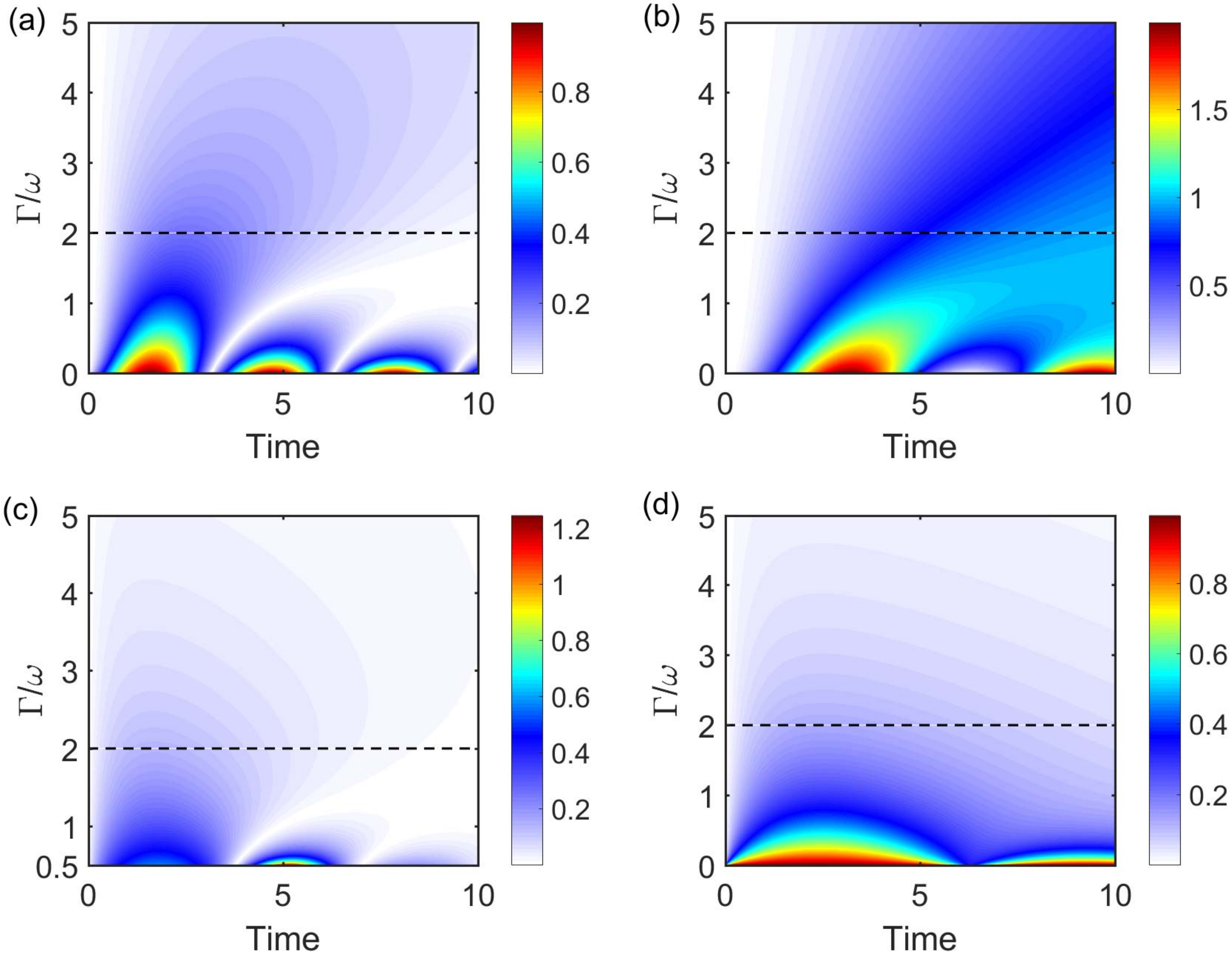}
	\caption{(Color online). (a) The absolute error of $\bar{y}(t;\Gamma|eq)$. (b) The absolute error of $\bar{w}(t;\Gamma|eq)$. (c) The relative error of $\bar{y}(t;\Gamma|eq)$. (d) The relative error of $\bar{w}(t;\Gamma|eq)$. The dashed line in each figure denotes the critical damping coefficient $\Gamma_c/\omega = 2$. These errors exhibit an oscillating behavior in time in the low-damping regime ($\Gamma < \Gamma_c$) while gradually reaching zero in the high-damping regime ($\Gamma > \Gamma_c$) as the damping coefficient increases.
	}
	\label{Fig:2Dplot_Eq}
\end{figure*}

\begin{figure*}[tbh]
	\includegraphics[scale=0.45,angle=0]{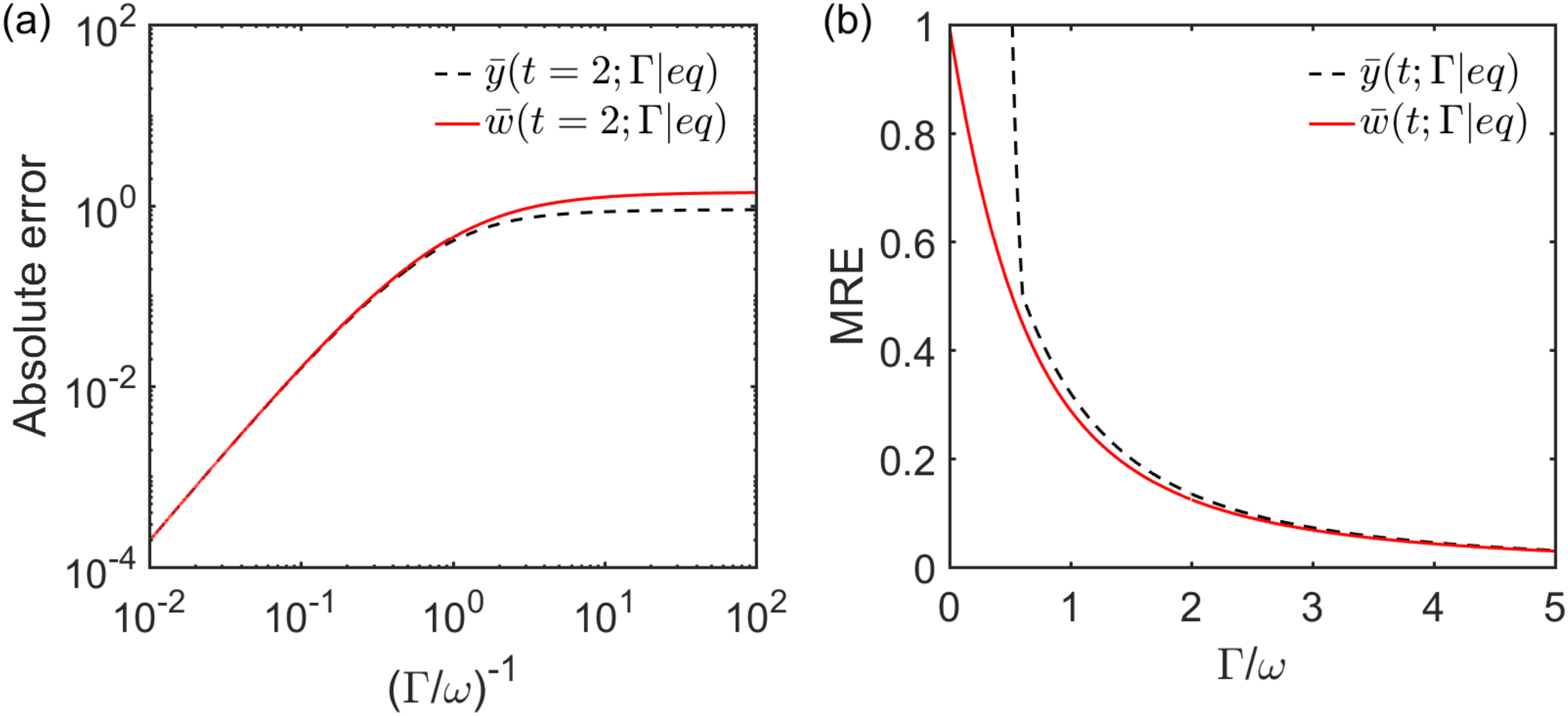}
	\caption{(Color online). (a) The absolute errors of $\bar{y}(t;\Gamma|eq)$ (black dashed curve) and $\bar{w}(t;\Gamma|eq)$ (red solid curve) at time $t=2$ as the function of $(\Gamma/\omega)^{-1}$ on the logarithmic scale. When $(\Gamma/\omega)^{-1}$ is small enough, the slopes of both the two curves are equal to 2. (b) The maximum relative errors of $\bar{y}(t;\Gamma|eq)$ (black dashed curve) and $\bar{w}(t;\Gamma|eq)$ (red solid curve) as the function of the damping coefficient $\Gamma$. They decrease, which means the OA is getting better, as the damping coefficient $\Gamma$ increases. When $\Gamma/\omega > 2$, MRE$\sk{\bar{y}}$ and MRE$\sk{\bar{w}}$ are almost the same. 
	}
	\label{Fig:MRE_Eq}
\end{figure*}

\begin{figure*}[t]
	\includegraphics[scale=0.42,angle=0]{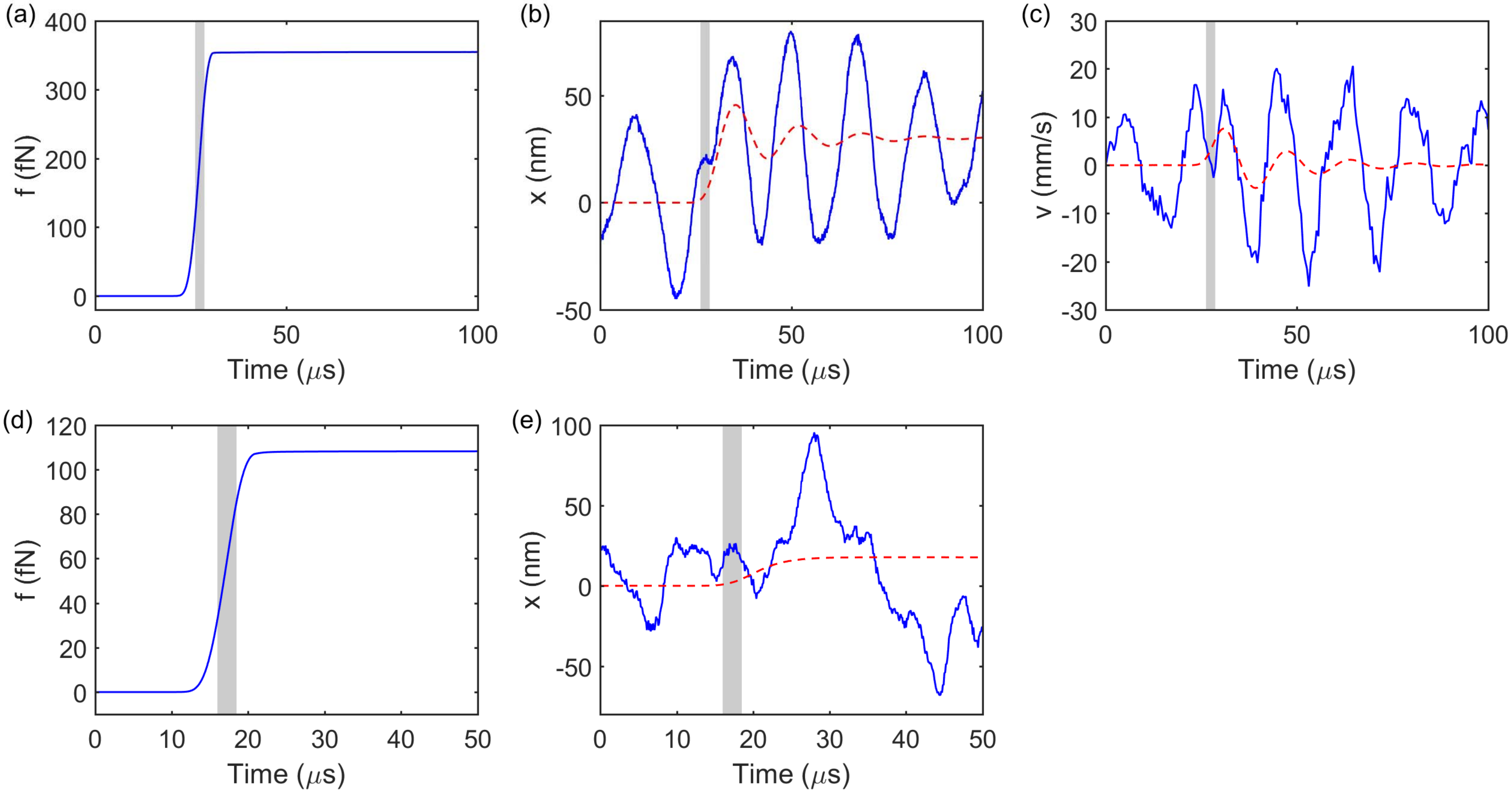}
	\caption{(Color online). Data of the force, the position, and the velocity as a function of time in the low-damping (50 Torr) and the high-damping regimes (760 Torr). (a) The force $f(t)$ exerted on the levitated sphere increased from zero to the final value of $354$ fN at 50 Torr. The shaded area indicates the time interval of interest. The time duration of the shaded area is $2.5\ \mu$s. During this period of time, the external driving force increases linearly in time, and the initial state obeys a nonequilibrium Gaussian distribution in the phase space. (b) The red dashed curve represents the averaged trajectory in the position space (over $10^6$ trajectories) at 50 Torr, and the blue solid curve is a single trajectory in the position space. (c) The red dashed curve represents the averaged trajectory in the velocity space (over $10^6$ trajectories) at 50 Torr, and the blue solid curve is a single trajectory in the velocity space. (d) The force $f(t)$ exerted on the levitated sphere increased from zero to the final value of $108$ fN at 760 Torr. The shaded area indicates the time interval of interest. The time duration of the shaded area is $2.6\ \mu$s. (e) The red dashed curve represents the averaged trajectory in the position space (over $10^6$ trajectories) at 760 Torr, and the blue solid curve is a single trajectory in the position space.
	}
	\label{Fig:Trajectory}
\end{figure*}

\begin{figure*}[t]
	\includegraphics[scale=0.42,angle=0]{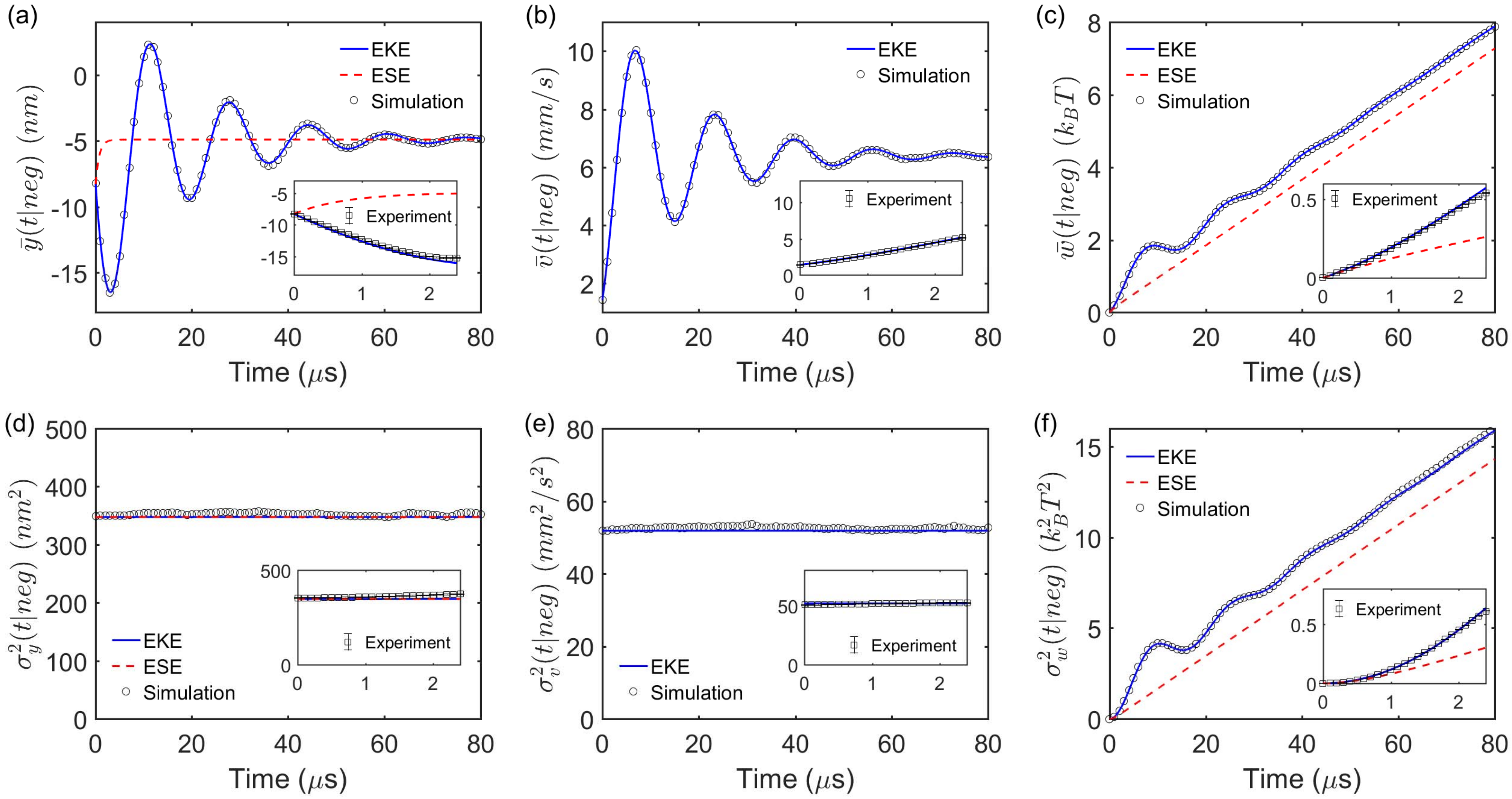}
	\caption{(Color online). Experimental data for the neg initial state in the low-damping regime (50 Torr). (a) Mean value of $y$. (b) Mean value of $v$. (c) Mean value of $w$. (d) Variance of $y$. (e) Variance of $v$. (f) Variance of $w$. The blue solid curves are solutions to the EKE, and the red dashed curves are solutions to the ESE. Results from the numerical simulation (circles) always agree well with the solutions to the EKE. The experimental data (squares) also agrees well with the solutions to the EKE. The error bars of the data represent the standard deviation of the measurements for 20 equal divisions of the initial state. Since the experimental duration is not long enough, we numerically simulate the linearly dragged process to illustrate the long-time behaviors of the levitated sphere.
	}
	\label{Fig:Underdamped_NEG}
\end{figure*}

\begin{figure*}[t]
	\includegraphics[scale=0.42,angle=0]{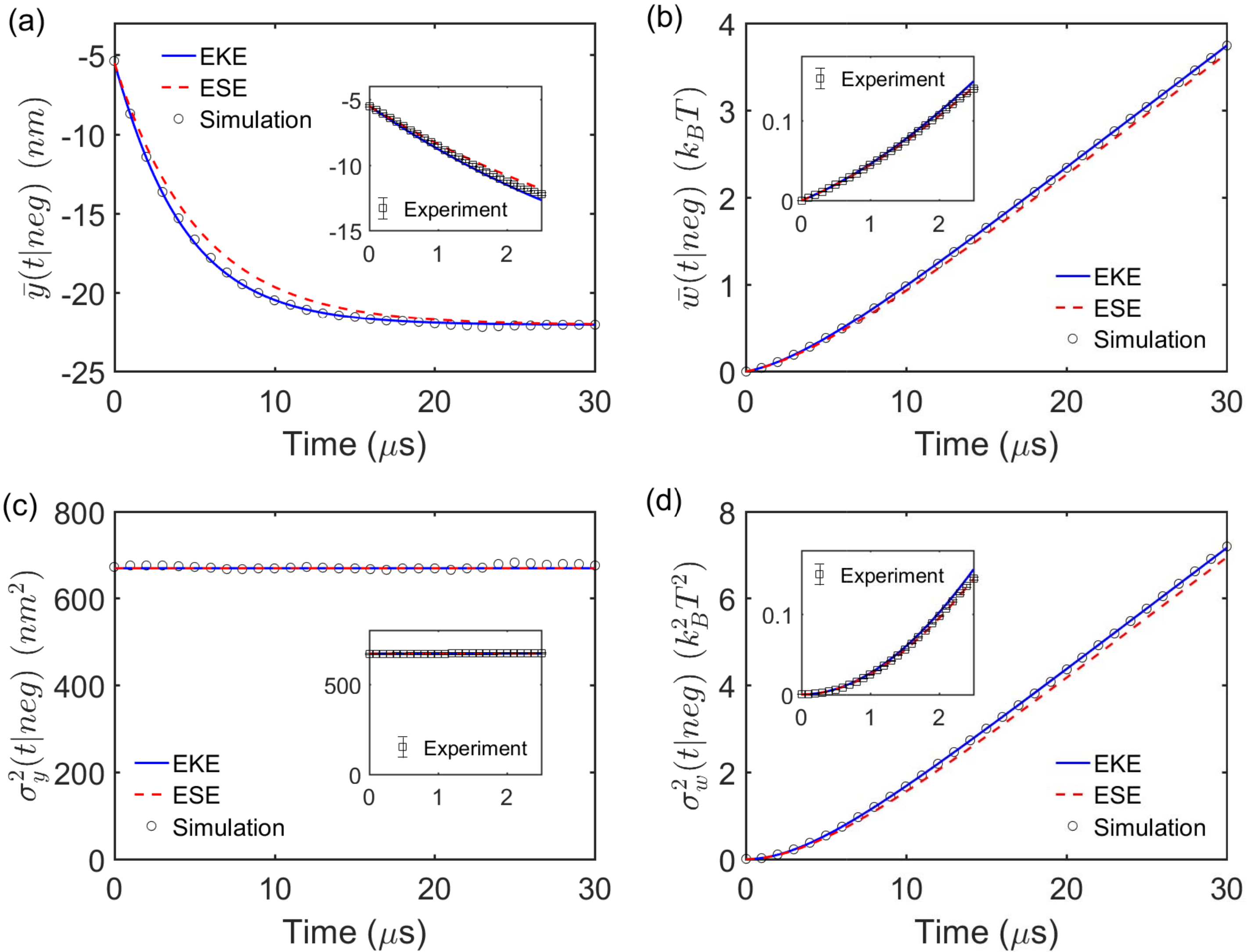}
	\caption{(Color online). Experimental data for the neg initial state in the high-damping regime (760 Torr). (a) Mean value of $y$. (b) Mean value of $w$. (c) Variance of $y$. (d) Variance of $w$. The solutions to the EKE (blue solid curves) are approaching the solutions to the ESE (red dashed curves) in this high-damping regime. Results from the numerical simulation (circles) always agree well with the solutions to the EKE. The experimental data (squares) also agrees well with the solutions to the EKE. The error bars of the data represent the standard deviation of the measurements for 20 equal divisions of the initial state.
	}
	\label{Fig:Overdamped_NEG}
\end{figure*}

\subsection{Miscellaneous discussions about the solution}

We analytically obtain the solution of $p(y,v,w,t|eq)$ [Eqs.~(\ref{HO:Gaussian}), (\ref{HO:Matrix}), (\ref{Solution:1}), (\ref{C12:Eq}), (\ref{Solution:2_Eq_U}), (\ref{Solution:3_Eq_U}), (\ref{D1D2E:Eq})], and verify the convergences of the moments of $p(y,v,w,t|eq)$ to their overdamped counterparts [Eqs.~(\ref{Solution:1_Eq_O}), (\ref{Solution_Eq_U:yvar}), (\ref{Solution:3_Eq_O})] in the overdamped limit. We notice that in Ref.~\cite{taniguchi2008inertial}, by utilizing the path integral approach, the work distribution of the same model is obtained analytically. In our current study, we also obtain an analytical expression of the work distribution. We cannot analytically prove the equivalence of the two distributions but we have numerically checked that they agree with each other. 
For this specific model, authors of Ref.~\cite{taniguchi2008inertial} also studied the convergence of the work distributions numerically. We would like to emphasize that our method is much simpler, and we are able to show the convergence analytically (see Sec. \ref{SubSec:EKE}).
Figure \ref{Fig:Analytical_Eq} shows the convergence behaviors of various moments [Eqs.~(\ref{Solution:1_Eq_O}) and (\ref{Solution:3_Eq_O})] with the equilibrium initial state. As we can see, with the increase of the damping coefficient $\Gamma$, the solutions to the EKE [blue (dark gray) curves] converge to the solutions to the ESE [green (light gray) curves].

It is important to notify that by comparing \eqnref{Solution_1:wbar} with \eqnref{Solution_Eq_U:wvar} and using the relation between $C_i$ and $D_i$ [\eqnref{D1D2E:Eq}], we find that
\beq
\bar{w}(t|eq) = \half \beta \sigma^2_w(t|eq).
\label{HO:FDT}
\eeq
This is a prediction of 
the Gaussian distribution \cite{mazonka1999exactly}, since for a Gaussian work distribution, only the first- and the second-order cumulants are nonzero, and all the higher-order cumulants are vanishing, i.e., $\bar{w} \equiv \Delta F + \half \beta \sigma^2_w$, and the free energy difference ($\Delta F$) in this model is equal to zero. The proportionality between $\bar{w}(t|eq)$ and $\sigma^2_w(t|eq)$ can be seen from \figref{(b)}{Fig:Analytical_Eq} and \figref{(d)}{Fig:Analytical_Eq}. Similarly, we find the following relations for this specific model:
\beqa
\bar{y}(t|eq) = \beta c_{yw}(t|eq),
\label{HO:FDT_yw}\\
\bar{v}(t|eq) = \beta c_{vw}(t|eq).
\label{HO:FDT_vw}
\eeqa
The proportionality between $\bar{y}(t|eq)$ and $c_{yw}(t|eq)$ 
can be seen from \figref{(a)}{Fig:Analytical_Eq} and \figref{(c)}{Fig:Analytical_Eq}. 

Using \eqnref{HO:FDT}, the Crooks fluctuation theorem \cite{crooks1999entropy}
\beq
\frac{p_{{\rm F}}(w)}{p_{{\rm R}}(-w)} = \ee^{\beta\rk{w - \Delta F}},
\label{CFT}
\eeq
and the Jarzynski equality \cite{jarzynski1997nonequilibrium} $\left\langle \ee^{-\beta\rk{w - \Delta F}} \right\rangle  = 1$ can be confirmed easily. Here ``F'' denotes the forward process, ``R'' the reverse process, and the work distribution of the reverse process $p_{{\rm R}}(w)$ can be calculated by replacing the shifting speed of the potential $u$ with $-u$ in $p_{{\rm F}}(w)$. Please note that the Crooks fluctuation theorem and the Jarzynski equality are valid for both the solutions to the EKE and the ESE, even though the two solutions are different.

In the low-damping regime ($\Gamma < \Gamma_c$), $\mu_{1,2}$ become complex numbers:
\beq
\mu_{1,2} = -\frac{\Gamma}{2} \pm \ii \Omega,
\label{HO:MuComplex}
\eeq
where
\beq
\Omega = \sqrt{\omega^2 - \rk{\frac{\Gamma}{2}}^2},\ \omega > \frac{\Gamma}{2}.
\label{HO:GammaOmega}
\eeq
Equations (\ref{Solution_1:wbar}) and (\ref{Solution_Eq_U:wvar}) can be rewritten in terms of trigonometric functions \cite{baule2009steady, funo2018path}
\beqa
\bar{w}(t|eq) = & & \frac{m u^2}{\omega^2} \left[ \omega^2 \Gamma t + \rk{\Gamma^2 - \omega^2}\rk{\ee^{-\frac{\Gamma}{2}t} \cos \Omega t - 1} \right. \nonumber\\
& & + \left. \frac{\Gamma}{2 \Omega}\rk{\Gamma^2 - 3 \omega^2}\ee^{-\frac{\Gamma}{2}t} \sin \Omega t \right],
\label{Solution:wbar_Complex_Eq}\\
\sigma^2_w(t|eq) = & & \frac{2 m u^2}{\beta \omega^2} \left[ \omega^2 \Gamma t + \rk{\Gamma^2 - \omega^2}\rk{\ee^{-\frac{\Gamma}{2}t} \cos \Omega t - 1} \right. \nonumber\\
& & + \left. \frac{\Gamma}{2 \Omega}\rk{\Gamma^2 - 3 \omega^2}\ee^{-\frac{\Gamma}{2}t} \sin \Omega t \right].
\label{Solution:wvar_Complex_Eq}
\eeqa
The trigonometric functions in Eqs.~(\ref{Solution:wbar_Complex_Eq}) and (\ref{Solution:wvar_Complex_Eq}) imply oscillation behaviors of $\bar{w}(t|eq)$ and $\sigma^2_w(t|eq)$. It can be seen from \figref{}{Fig:Analytical_Eq} that $\bar{w}(t|eq)$ and $\sigma^2_w(t|eq)$ do oscillate when $\Gamma < \Gamma_c$, which is a character of the underdamped stochastic thermodynamics. Similar analysis can be applied to $\bar{y}(t|eq)$, $\bar{v}(t|eq)$, $c_{yw}(t|eq)$, and $c_{vw}(t|eq)$.

In addition, in the long time limit $t \rightarrow +\infty$, since the real parts of $\mu_{1,2}$ and $\lambda_{1,2,3}$ are all negative, the exponential terms in the solutions of the moments will vanish. Hence the asymptotic behaviors of the solutions to the EKE and the ESE are the same in the long time limit, regardless of the initial distribution \cite{mazonka1999exactly, taniguchi2008inertial}.

Here we would like to quantify the accuracy of the OA as a function of the damping coefficient $\Gamma$. Now we have two sets of solutions. One is from the EKE, while the other one is from the ESE. Taking $\bar{y}(t;\Gamma|eq)$ and $\bar{w}(t;\Gamma|eq)$ for examples, we show the absolute errors [\figref{(a)}{Fig:2Dplot_Eq} and \figref{(b)}{Fig:2Dplot_Eq}] and the relative errors [\figref{(c)}{Fig:2Dplot_Eq} and \figref{(d)}{Fig:2Dplot_Eq}] of $\bar{y}(t;\Gamma|eq)$ and $\bar{w}(t;\Gamma|eq)$, respectively, as the function of time $t$ and the damping coefficient $\Gamma$. The relative errors [\figref{(c)}{Fig:2Dplot_Eq} and \figref{(d)}{Fig:2Dplot_Eq}] vanish with the increase of $\Gamma$ and $t$. We can see the oscillating behaviors in the low-damping regime ($\Gamma < \Gamma_c$) in each figure. In \figref{(a)}{Fig:MRE_Eq}, we plot the absolute errors of $\bar{y}(t;\Gamma|eq)$ and $\bar{w}(t;\Gamma|eq)$ at time $t=2$ as the function of $(\Gamma/\omega)^{-1}$ on the logarithmic scale. When $(\Gamma/\omega)^{-1}$ is small enough, the slopes of both the two curves are equal to 2, which means the leading order of the absolute errors is proportional to $\Gamma^{-2}$. This is consistent with the prediction of \eqnref{Eqn:Expansion}. We use the quantity of the maximum relative error (MRE) during the time interval [$0,+\infty$) to characterize the accuracy of the OA as a function of $\Gamma$:
\begin{subequations}
\beqa
& & {\rm MRE}[\bar{y}] := {\rm max}_{t \in [0,+\infty)} \left| \frac{\bar{y}^{{\rm EKE}}(t;\Gamma) - \bar{y}^{{\rm ESE}}(t;\Gamma)}{\bar{y}^{{\rm EKE}}(t;\Gamma)} \right|,
\label{MREy}\\
& & {\rm MRE}[\bar{w}] := {\rm max}_{t \in [0,+\infty)} \left| \frac{\bar{w}^{{\rm EKE}}(t;\Gamma) - \bar{w}^{{\rm ESE}}(t;\Gamma)}{\bar{w}^{{\rm EKE}}(t;\Gamma)} \right|,
\label{MREw}
\eeqa
\label{MRE}
\end{subequations}
where the superscript ``EKE'' denotes the solutions to the EKE, and ``ESE'' the solutions to the ESE. In \figref{(b)}{Fig:MRE_Eq}, we plot the MREs of $\bar{y}(t;\Gamma|eq)$ and $\bar{w}(t;\Gamma|eq)$. Both MREs [\eqnref{MRE}] decrease smoothly across the boundary of the two regimes, which means the OA is getting better, as the damping coefficient $\Gamma$ increases. We can find that the OA leads to a deviation of 30\% from the exact solution when $\Gamma = \omega$, 13\% for $\Gamma = \Gamma_c = 2\omega$, 10\% for $\Gamma = 2.5 \omega$, and 3\% for $\Gamma = 5 \omega$. If we regard a deviation of less than 10\% as a good approximation, $\Gamma \geq 2.5 \omega$ is required for the validity of the OA. 



\section{\label{Sec:Experiment}Experimental demonstration}

We perform an experiment and compare the data with the theoretical results obtained in Sec. \ref{Sec:LDHO}. We find that the experimental data always agrees with the results from solving the EKE but deviates from the solutions to the ESE in the low-damping regime. Our experimental data demonstrates the convergence of the distributions of work and the position as the damping coefficient increases.

Our experiment is carried out using a silica nanosphere levitated in air by an optical tweezer at room temperature 296 K. The optical tweezer is generated by a focused 1550-nm laser beam. A series of 532-nm laser pulses exert a time-dependent transverse scattering force on the particle to drive it out of equilibrium \cite{hoang2018experimental}. The optical trap can be regarded as a harmonic potential, and the transverse force shifts the center of the potential linearly in time. The potential can be expressed as
\beq
\mathcal{U}(x,t) = \half m \omega^2 \sk{x - \frac{f(t)}{m \omega^2}}^2,
\label{Ex:Potential}
\eeq
where $f(t)$ is the transverse force that changes linearly in time from $t=0$ to $t=t_f$. Compared with \eqnref{HO:Potential}, the dragging speed $u$ satisfies
\beq
f(t) = mu \omega^2 t.
\label{Ex:DraggingSpeed}
\eeq
The trajectory work is calculated as \cite{sekimoto2010stochastic}
\beq
w = \int_{f(0)}^{f(t_f)} \pd{\mathcal{U}}{f} \dd f = -\int_{f(0)}^{f(t_f)} \rk{x - \frac{f}{m \omega^2}} \dd f
\label{Ex:Work}
\eeq

Both the position and the instantaneous velocity of the levitated nanosphere are measured by our ultrasensitive detector \cite{li2010measurement}. The same experimental procedure is repeated for over one million times. About $10^{10}$ position data are recorded with a 10 MHz acquisition rate to provide sufficient statistics. 

The damping coefficient $\Gamma$ is controlled by tuning the air pressure \cite{hoang2016electron}. We carry out our experiment under two kinds of pressures: 50 Torr and 760 Torr (1 atm), respectively. The case of 50 Torr corresponds to the low-damping regime, and the case of 760 Torr corresponds to the high-damping regime. At 50 Torr, we trap a 212-nm-radius silica nanosphere with a trapping frequency $61.5 \pm 0.3$ (2$\pi \cdot$ kHz). The velocity relaxation time $\tau_p = \Gamma^{-1} = 8.8\ \mu$s
. At 760 Torr, we trap a 145-nm-radius silica nanosphere with a trapping frequency $78 \pm 3$ (2$\pi \cdot$ kHz). The velocity relaxation time $\tau_p = \Gamma^{-1} = 0.79\ \mu$s
. Figure \ref{Fig:Trajectory} shows the force and the trajectories in the position and the velocity spaces in both regimes (pressures). We focus our attention on the middle part of the driving protocol [see the shaded areas in \figref{(a)}{Fig:Trajectory} and \figref{(d)}{Fig:Trajectory}], which can be described by the linearly dragged harmonic oscillator model introduced in Sec.~\ref{Sec:LDHO}. At 50 Torr [\figref{(a)}{Fig:Trajectory}], the time duration of the shaded area is $2.5\ \mu$s. The system is in the low-damping regime (
$\Gamma/\omega = 0.30 < 2$), and the particle is driven out of equilibrium [witnessed by the oscillations of the red dashed curves in \figref{(b)}{Fig:Trajectory} and \figref{(c)}{Fig:Trajectory}]. At 760 Torr [\figref{(d)}{Fig:Trajectory}], the time duration of the shaded area is $2.6\ \mu$s. The system is in the high-damping regime (
$\Gamma/\omega = 2.57 > 2$), and the particle is driven out of equilibrium [witnessed by the red dashed curve in \figref{(e)}{Fig:Trajectory} without oscillation]. The velocity is not included in the high-damping regime because it is usually assumed that the distribution of the velocity will relax to the thermal equilibrium distribution instantaneously.

The initial state of the driving process in the shaded area is not the equilibrium state, but a center-shifted nonequilibrium Gaussian (neg) state. During the driving process, in comparison with the equilibrium initial state [\eqnref{InitialState:eq}], the evolutions of $\sigma^2_y$, $\sigma^2_v$, $c_{yv}$, $\sigma^2_w$, $c_{yw}$, and $c_{vw}$ remain unchanged [Eqs.~(\ref{Solution:2_Eq_U}), (\ref{Solution:3_Eq_U}), and (\ref{D1D2E:Eq})]. In addition, variances of both distributions in the position space and the velocity space remain a constant [\eqnref{Solution:2_Eq_U}]. However, the evolutions of the mean values ($\bar{y}$, $\bar{v}$, and $\bar{w}$) are shifted. By substituting the initial condition of the neg state
\beq
\bar{y}(0|neg) = y_0,\ \bar{v}(0|neg) = v_0
\label{InitialState:NEG}
\eeq
into \eqnref{Solution:1}, we obtain
\beqa
C_1 &=& \frac{v_0-u-\rk{y_0 + \frac{u\Gamma}{\omega^2}}\mu_2}{\mu_1 - \mu_2} \rightarrow y_0 + \frac{u\Gamma}{\omega^2},\nonumber\\
C_2 &=& \frac{-v_0 +u+\rk{y_0 + \frac{u\Gamma}{\omega^2}}\mu_1}{\mu_1 - \mu_2} \rightarrow -\frac{v_0}{\Gamma},
\label{C12:NEG}
\eeqa
and
\begin{subequations}
\beqa
& & \bar{y}(t|neg) \rightarrow \rk{y_0 +\frac{u\Gamma}{\omega^2}} \ee^{-\frac{\omega^2}{\Gamma} t}-\frac{u\Gamma}{\omega^2},
\label{Solution_NEG_O:ybar}\\
& & \bar{w}(t|neg) \rightarrow m u^2 \Gamma t + m u \Gamma \rk{y_0 + \frac{u\Gamma}{\omega^2}} \rk{\ee^{-\frac{\omega^2}{\Gamma} t}-1}.
\label{Solution_NEG_O:wbar}
\eeqa
\label{Solution:1_NEG_O}
\end{subequations}
Please note that the r.h.s. of \eqnref{Solution:1_NEG_O} is identical to the solutions to the ESE obtained in Ref.~\cite{mazonka1999exactly}. When $y_0 = v_0 = 0$, \eqnref{Solution:1_NEG_O} reproduces \eqnref{Solution:1_Eq_O}.

We compare the analytical solutions with the experimental data in both low-damping (50 Torr) and high-damping (760 Torr) regimes. As shown in \figref{}{Fig:Underdamped_NEG} (50 Torr) and \figref{}{Fig:Overdamped_NEG} (760 Torr), 
both the experimental data and simulation results show good agreements with the analytical solutions to the EKE (since the experimental duration is not long enough, by using the complete LE, we numerically simulate the linearly dragged process for a long time until the distribution reaches a steady distribution). By comparing \figref{}{Fig:Underdamped_NEG} and \figref{}{Fig:Overdamped_NEG}, we find that in the low-damping regime (\figref{}{Fig:Underdamped_NEG}), the solutions to the EKE differ significantly from those to the ESE, which indicates that the OA breaks down in this regime. For example, the OA leads to a deviation of 62.4\% from the exact solution of $\bar{w}(t|neg)$, and 66.8\% for $\sigma^2_w(t|neg)$. However, the solutions to the EKE almost overlap with those to the ESE in \figref{}{Fig:Overdamped_NEG}, which indicates the OA is a good approximation in the case of high damping coefficient. For example, the OA leads to a deviation of 6.97\% from the exact solution of $\bar{y}(t|neg)$, 6.17\% for $\bar{w}(t|neg)$, and 8.68\% for $\sigma^2_w(t|neg)$. One can expect that with the increase of the damping coefficient, the convergence of the blue solid and the red dashed curves will be even better.


\section{\label{Sec:Conclusion}Conclusion}

In summary, in this article, we quantitatively study the validity and breakdown of the OA for the work distribution in stochastic thermodynamics both theoretically and experimentally. Firstly, we consider the general case starting from the extended Fokker-Planck equations including work, and confirm the convergence of the EKE to the ESE using a multiple timescale expansion approach. 
Thus we show the convergence behavior of the work distributions explicitly in the overdamped limit. We find that the leading order of the deviation of the OA from the exact solution is proportional to $\Gamma^{-2}$ [\eqnref{Eqn:Expansion}]. 
Then, as an exactly solvable example, we calculate the joint probability distribution of the position, the velocity, and work for a linearly dragged harmonic oscillator. The convergence of the solution of the EKE to the solution of the ESE in the overdamped limit is verified. 
In this specific model, we use the quantity of the MRE [\eqnref{MRE}] to characterize the accuracy of the OA as a function of the damping coefficient, and find the OA leads to a deviation of 10\% from the exact solution when the damping coefficient is 2.5 times of the trapping frequency, and 3\% for 5 times of the trapping frequency. In addition, we experimentally demonstrate that the data of the work distribution of a levitated silica nanosphere always agrees with the solution to the EKE, thus it agrees with the OA in the overdamped limit, but deviates from the OA in the low-damping case. 
Our work 
fills a gap between the stochastic thermodynamics based on the complete LE (the KE) and the overdamped LE (the SE), and deepens our understanding of the OA 
in stochastic thermodynamics.

\begin{acknowledgments}
H. T. Q. acknowledges support from the National Science Foundation of China under Grants No. 11775001 and No. 11534002, and the Recruitment Program of Global Youth Experts of China. T. L. acknowledges support from the National Science Foundation under Award No. 1555035-PHY and the Office of Naval Research under Award No. N00014-18-1-2371. 
\end{acknowledgments}

\appendix

\setcounter{figure}{0}
\renewcommand{\thefigure}{A\arabic{figure}}

\section{\label{A:OverdampedEqns}Derivation of Eqs.~(\ref{Eqn:OLEW0}) and (\ref{Eqn:OFPW0})}

In the overdamped LE [\eqnref{Eqn:OLE0}], according to the definition of the work functional along individual trajectories [\eqnref{WorkDefinition}], we can rewrite the Stratonovich product into the It\^o product,
\beqa
\dj w = & & \pd{\mathcal{U}}{t} \dd t + \mathcal{F}_{nc} \circ \dd x \nonumber\\
= & & \pd{\mathcal{U}}{t} \dd t + \mathcal{F}_{nc} \cdot \dd x + \half \dd \mathcal{F}_{nc} \dd x \nonumber\\
= & & \pd{\mathcal{U}}{t} \dd t + \mathcal{F}_{nc} \cdot \dd x + \Gamma^{-1} \frac{1}{\beta m} \mathcal{F}_{nc}' \dd t \nonumber\\
= & & \left[ \pd{\mathcal{U}}{t} + \Gamma^{-1} \frac{1}{m} \left( \mathcal{F} \mathcal{F}_{nc} + \frac{1}{\beta} \mathcal{F}_{nc}' \right) \right] \dd t \nonumber\\
& & + \sqrt{\frac{2 \Gamma^{-1}}{\beta m}} \mathcal{F}_{nc} \cdot \dd \mathcal{B}_t.
\label{Der:OLEW0}
\eeqa
This is \eqnref{Eqn:OLEW0}.

By combining \eqnref{Der:OLEW0} with \eqnref{Eqn:OLE0} and defining
\beq
\bm{Y} = \left(
\begin{array}{c}
x \\ w
\end{array} \right), \nabla_{\bm{Y}} = \left(
\begin{array}{c}
\pd{}{x} \\ \pd{}{w}
\end{array} \right),
\eeq
we obtain the Langevin equation of $\bm{Y}$:
\beq
\dd \bm{Y} = \bm{\Pi} \dd t + \bm{\Lambda} \dd \bm{B}_t,
\label{LE:Y}
\eeq
where the drift coefficient $\bm{\Pi}$ and the noise coefficient $\bm{\Lambda}$ are
\beq
\bm{\Pi} = \left(
\begin{array}{c}
\Gamma^{-1} \frac{1}{m} \mathcal{F} \\ \pd{\mathcal{U}}{t} + \Gamma^{-1} \frac{1}{m} \left( \mathcal{F} \mathcal{F}_{nc} + \frac{1}{\beta} \mathcal{F}_{nc}' \right)
\end{array} \right),
\eeq
\beq
\bm{\Lambda} = \sqrt{\frac{2 \Gamma^{-1}}{\beta m}} \left(
\begin{array}{cc}
1 & 0 \\
\mathcal{F}_{nc} & 0
\end{array}
\right),
\eeq
and
\beq
\dd \bm{B}_t = \left(
\begin{array}{c}
\dd \mathcal{B}_t \\ \dd \tilde{\mathcal{B}}_t
\end{array} \right)
\eeq
consists of two independent noises.

Then the extended Fokker-Planck equation of $x$ and $w$ can be written as
\begin{widetext}
\beqa
\pd{}{t}\rho(x,w;t) &=& - \nabla_{\bm{Y}} \cdot \rk{\bm{\Pi} \rho} + \nabla_{\bm{Y}} \cdot \sk{\nabla_{\bm{Y}} \cdot \rk{\bm{D} \rho}} \nonumber\\
&=& - \frac{\Gamma^{-1}}{m} \pd{}{x} \rk{\mathcal{F} \rho} -  \sk{\pd{\mathcal{U}}{t} + \frac{\Gamma^{-1}}{m} \left( \mathcal{F} \mathcal{F}_{nc} + \frac{1}{\beta}\mathcal{F}_{nc}' \right)} \pd{\rho}{w} + \frac{\Gamma^{-1}}{\beta m} \sk{ \frac{\partial^2 \rho}{\partial x^2} + 2 \pd{}{x} \rk{\mathcal{F}_{nc} \pd{\rho}{w}} + \mathcal{F}_{nc}^2 \frac{\partial^2 \rho}{\partial w^2}} \nonumber\\
&=& \frac{\Gamma^{-1}}{m} \bk{ \pd{}{x} \left(\frac{1}{\beta} \pd{\rho}{x} -\mathcal{F} \rho \right) - \mathcal{F} \mathcal{F}_{nc} \pd{\rho}{w} + \frac{1}{\beta} \sk{ \mathcal{F}_{nc} \pd{}{w} \pd{\rho}{x} + \pd{}{x} \rk{\mathcal{F}_{nc} \pd{\rho}{w}} + \mathcal{F}_{nc}^2 \frac{\partial^2 \rho}{\partial w^2}}} - \pd{\mathcal{U}}{t} \pd{\rho}{w} \nonumber\\
&=& \Gamma^{-1} \frac{1}{m} \rk{\pd{}{x} + \mathcal{F}_{nc} \pd{}{w}} \rk{\frac{1}{\beta} \pd{}{x} - \mathcal{F} + \frac{1}{\beta} \mathcal{F}_{nc} \pd{}{w}} \rho(x,w;t) - \pd{\mathcal{U}}{t} \pd{\rho}{w},
\label{Der:OFPW0}
\eeqa
\end{widetext}
where
\beqa
\bm{D} = \half \bm{\Lambda} \bm{\Lambda}^T = \frac{\Gamma^{-1}}{\beta m} \left(
\begin{array}{cc}
1 & \mathcal{F}_{nc} \\
\mathcal{F}_{nc} & \mathcal{F}_{nc}^2
\end{array}
\right),
\eeqa
is the diffusion coefficient. This is the ESE [\eqnref{Eqn:OFPW0}].

\section{\label{A:MTSA}Proof of the convergence of \eqnref{Eqn:UFPW0} to \eqnref{Eqn:OFPW0}: a multiple timescale approach}

\subsection{Dimensionless extended Kramers equation and extended Smoluchowski equation}

For convenience, we introduce the following dimensionless variables:
\beqa
\begin{split}
& \tau=t \frac{v_T}{d},\ X=\frac{x}{d},\ V=\frac{v}{v_T},\ B_\tau=\mathcal{B}_t \sqrt{\frac{v_T}{d}},
\\& F=\mathcal{F}\frac{d}{mv_T^2},\ f_{nc}=\mathcal{F}_{nc}\frac{d}{mv_T^2},\ U=\frac{\mathcal{U}}{mv_T^2},
\\& W=\frac{w}{mv_T^2},\ \xi = \frac{\gamma}{m}\frac{d}{v_T},
\end{split}
\label{Dimensionless}
\eeqa
where $d$ is the particle size, $v_T=\sqrt{\bz T/m}$ is the thermal velocity, and $\xi$ is the dimensionless damping coefficient. Then the EKE [\eqnref{Eqn:UFPW0}] can be rewritten in the dimensionless form
\begin{widetext}
\beq
\pd{}{V} \left( V+ \pd{}{V}\right) p(X,V,W;\tau) =
\xi^{-1} \left[ \pd{}{\tau} + V\pd{}{X}+ F\pd{}{V} + \left( \pd{U}{\tau} + f_{nc}V \right) \pd{}{W} \right] p(X,V,W;\tau),
\label{Eqn:UFPW}
\eeq
\end{widetext}
where $p(X,V,W;\tau)$ is the joint probability distribution of the position, the velocity, and work at time $\tau$ associated with the EKE [\eqnref{Eqn:UFPW}]. 

The ESE [\eqnref{Eqn:OFPW0}] can be rewritten in the dimensionless form
\begin{widetext}
\beq
\pd{}{\tau}\rho(X,W;\tau) = \xi^{-1} \rk{\pd{}{X} + f_{nc}\pd{}{W}} \rk{\pd{}{X} - F + f_{nc}\pd{}{W}} \rho(X,W;\tau) - \pd{U}{\tau} \pd{\rho}{W},
\label{Eqn:OFPW}
\eeq
\end{widetext}
where $\rho(X,W;\tau)$ is the joint probability distribution of the position and work at time $\tau$ associated with the ESE [\eqnref{Eqn:OFPW}].

\subsection{Proof of the convergence of \eqnref{Eqn:UFPW0} to \eqnref{Eqn:OFPW0}}

In this subsection, we rigorously prove the convergence of \eqnref{Eqn:UFPW} to \eqnref{Eqn:OFPW} in the overdamped limit ($\xi^{-1}\rightarrow 0$), thus we prove the convergence of the work distribution associated with the EKE to that associated with the ESE. In this way, we verify the validity of the OA for the work distribution in stochastic thermodynamics in the overdamped limit.

We utilize the multiple timescale approach \cite{bocquet1997high, wang2016entropy, celani2012anomalous, bo2016multiple}. 
Time is separated by
\beq
\tau_0 = \tau;\ \tau_1 = \xi^{-1} \tau;\ \tau_2 = \xi^{-2} \tau;\ \cdots
\label{Timescales}
\eeq
so that the time derivative is replaced by the summation of time derivatives at different scales,
\beq
\pd{}{\tau} \rightarrow \pd{}{\tau_0} + \xi^{-1} \pd{}{\tau_1} + \xi^{-2} \pd{}{\tau_2} + \cdots,
\label{TimeSeparation}
\eeq
and the probability distribution can also be expanded in the orders of $\xi^{-1}$,
\beqa
p(X,V,W;\tau) = & & p^{(0)}(X,V,W;\tau_0,\tau_1,\cdots) + \nonumber \\
& & \xi^{-1} p^{(1)}(X,V,W;\tau_0,\tau_1,\cdots) + \cdots.
\label{pdfSeparation}
\eeqa
The external forces are assumed to vary at the slow timescales only, i.e.,
\beq
\pd{U}{\tau_0} = \pd{F}{\tau_0} = \pd{f_{nc}}{\tau_0} = 0.
\label{SlowPotential}
\eeq
We define an operator $\mathcal{L} = \pd{}{V} \rk{V + \pd{}{V}}$. By substituting \eqnref{pdfSeparation} into \eqnref{Eqn:UFPW} and identifying terms of the same order of $\xi^{-1}$, we obtain
\begin{widetext}
\beqa
\begin{split}
& \mathcal{L} p^{(0)} = 0;\\
& \mathcal{L} p^{(1)} = \sk{\pd{}{\tau_0} + V\pd{}{X} + F\pd{}{V} +  V f_{nc} \pd{}{W}} p^{(0)};\\
& \mathcal{L} p^{(2)} = \sk{\pd{}{\tau_0} + V\pd{}{X} + F\pd{}{V} + V f_{nc} \pd{}{W}} p^{(1)} + \sk{\pd{}{\tau_1} + \pd{U}{\tau_1} \pd{}{W}} p^{(0)};\\
& \cdots
\end{split}
\label{Eqn:Identification}
\eeqa
\end{widetext}

The zeroth-order equation implies a Maxwellian distribution in the velocity space, and the velocity can be separated from the position and work:
\beq
p^{(0)}(X,V,W;\tau_0,\tau_1,\cdots) = \Phi(X,W;\tau_0,\tau_1,\cdots)\ee ^{-\half V^2},
\label{Sol:Zeroorder}
\eeq
where the function $\Phi$ is to be determined. Then the first-order equation follows
\beqa
\mathcal{L} p^{(1)} = & & \pd{\Phi}{\tau_0} \ee ^{-\half V^2} \nonumber\\
& & + V \rk{\pd{\Phi}{X} - F\Phi + f_{nc}\pd{\Phi}{W}} \ee ^{-\half V^2}.
\label{Eqn:Firstorder}
\eeqa
By integrating on both sides of \eqnref{Eqn:Firstorder} over $V$, we obtain
\beq
\pd{\Phi}{\tau_0} = 0.
\label{Eqn:SolubilityConditionI}
\eeq
This is called the ``solubility condition'' \cite{bocquet1997high}. By combining \eqnref{Eqn:Firstorder} and \eqnref{Eqn:SolubilityConditionI}, we obtain the first-order correction to the probability distribution,
\beqa
p^{(1)}(X,V,W;\tau_0,\cdots) = & & - V \rk{\pd{\Phi}{X} - F\Phi + f_{nc}\pd{\Phi}{W}} \ee ^{-\half V^2} \nonumber\\
& & + \Psi(X,W;\tau_0,\tau_1,\cdots)\ee ^{-\half V^2},
\label{Sol:Firstorder}
\eeqa
where the function $\Psi$ is to be determined. Substituting Eqs.~(\ref{Sol:Zeroorder}) and (\ref{Sol:Firstorder}) into \eqnref{Eqn:Identification}, after some algebra, the second-order equation can be written as
\begin{widetext}
\beqa
\mathcal{L} p^{(2)} = & & \sk{\pd{\Psi}{\tau_0} + \pd{\Phi}{\tau_1} + \pd{U}{\tau_1} \pd{\Phi}{W} - \rk{\pd{}{X} + f_{nc}\pd{}{W}} \rk{\pd{\Phi}{X} - F\Phi + f_{nc}\pd{\Phi}{W}}} \ee ^{-\half V^2} \nonumber\\
& & + V \sk{\pd{\Psi}{X} - F\Psi + f_{nc}\pd{\Psi}{W}}  \ee ^{-\half V^2} \nonumber\\
& & + \rk{\pd{}{X} - F + f_{nc}\pd{}{W}} \rk{\pd{\Phi}{X} - F\Phi + f_{nc}\pd{\Phi}{W}} \rk{1-V^2} \ee ^{-\half V^2}.
\label{Eqn:Secondorder}
\eeqa
\end{widetext}
Similarly, 
by integrating on both sides of \eqnref{Eqn:Secondorder} over $V$, we obtain the second ``solubility condition''
\beqa
\pd{\Psi}{\tau_0} = & & - \pd{\Phi}{\tau_1} - \pd{U}{\tau_1} \pd{\Phi}{W} \nonumber\\
& & + \rk{\pd{}{X} + f_{nc}\pd{}{W}} \rk{\pd{\Phi}{X} - F\Phi + f_{nc}\pd{\Phi}{W}}.
\label{Eqn:SolubilityConditionII}
\eeqa
The r.h.s. of this equation does not depend on $\tau_0$ because of \eqnref{Eqn:SolubilityConditionI}, and one must impose the condition
\beq
\pd{\Psi}{\tau_0} = 0
\label{Eqn:SolubilityConditionII2}
\eeq
to eliminate the secular divergence as $\tau_0$ grows to infinity \cite{bocquet1997high}. This leads to a closed equation for $\Phi$:
\beqa
\pd{\Phi}{\tau_1} = & & \rk{\pd{}{X} + f_{nc}\pd{}{W}} \rk{\pd{\Phi}{X} - F\Phi + f_{nc}\pd{\Phi}{W}} \nonumber\\
& & - \pd{U}{\tau_1} \pd{\Phi}{W}.
\label{Eqn:SolubilityConditionII3}
\eeqa
Please note that \eqnref{Eqn:SolubilityConditionII2} and \eqnref{Eqn:SolubilityConditionII3} are helpful in deriving \eqnref{Eqn:Expansion}.

By collecting Eqs.~(\ref{Sol:Zeroorder}) and (\ref{Sol:Firstorder}), the joint probability distribution can be written as (to the order of $\xi^{-2}$):
\beqa
p(X,V,W;\tau) = & & \ee ^{-\half V^2} \left[ \Phi - \xi^{-1} V\rk{\pd{\Phi}{X} - F\Phi + f_{nc}\pd{\Phi}{W}} \right. \nonumber\\
& & \left. + \xi^{-1} \Psi + \mathcal{O}\rk{\xi^{-2}} \right].
\label{Sol:XVWpdf}
\eeqa
The joint probability distribution of the position and work can be obtained by integrating on both sides of \eqnref{Sol:XVWpdf} over $V$,
\beqa
\tilde{p}(X,W;\tau) &=& \int p(X,V,W;\tau) \dd V \nonumber\\
&=& \sqrt{2\pi} \sk{\Phi + \xi^{-1} \Psi + \mathcal{O}\rk{\xi^{-2}}}.
\label{Sol:XWpdf}
\eeqa
We do the time derivative [\eqnref{TimeSeparation}] over \eqnref{Sol:XWpdf}, and the equation of motion for $\tilde{p}(X,W;\tau)$ can be written as
\begin{widetext}
\beqa
\pd{}{\tau}\tilde{p}(X,W;\tau) = & & \sqrt{2\pi} \sk{\pd{\Phi}{\tau_0} + \xi^{-1} \rk{\pd{\Psi}{\tau_0} + \pd{\Phi}{\tau_1}} + \mathcal{O}\rk{\xi^{-2}}} \nonumber\\
= & & \sqrt{2\pi} \bk{ \xi^{-1} \sk{ - \pd{U}{\tau_1} \pd{\Phi}{W} + \rk{\pd{}{X} + f_{nc}\pd{}{W}} \rk{\pd{\Phi}{X} - F\Phi + f_{nc}\pd{\Phi}{W}}} + \mathcal{O}\rk{\xi^{-2}}} \nonumber\\
= & & \xi^{-1} \rk{\pd{}{X} + f_{nc}\pd{}{W}} \rk{\pd{}{X} - F + f_{nc} \pd{}{W}} \tilde{p} - \pd{U}{\tau} \pd{\tilde{p}}{W} + \mathcal{O}\rk{\xi^{-2}},
\label{Eqn:OFPW2}
\eeqa
\end{widetext}
where we have used the two ``solubility conditions'' [Eqs.~(\ref{Eqn:SolubilityConditionI}) and (\ref{Eqn:SolubilityConditionII})]. It can be seen that \eqnref{Eqn:OFPW2} is exactly the same as \eqnref{Eqn:OFPW} when the higher-order corrections are ignored in the overdamped limit. Hence we have rigorously proved the convergence of \eqnref{Eqn:UFPW} to \eqnref{Eqn:OFPW}, namely, \eqnref{Eqn:UFPW0} to \eqnref{Eqn:OFPW0}, in the overdamped limit [see \eqnref{Eqn:OFPW02}].

\section{\label{A:EqInitial}Solutions of the second-order moments and their asymptotic behaviors}

Equations (\ref{Couple:2}) and (\ref{EOM:d}), (\ref{EOM:e}), and (\ref{EOM:g}) give the solutions of the moments about $y$ and $v$,
\begin{subequations}
\beqa
& & \sigma^2_y(t) = B_1 \ee^{\lambda_1 t} + B_2 \ee^{\lambda_2 t} + B_3 \ee^{\lambda_3 t} + \frac{1}{\beta m \omega^2},
\label{Solution_2:yvar}\\
& & \sigma^2_v(t) = F_1 B_1 \ee^{\lambda_1 t} + F_2 B_2 \ee^{\lambda_2 t} + F_3 B_3 \ee^{\lambda_3 t} + \frac{1}{\beta m},
\label{Solution_2:vvar}\\
& & c_{yv}(t) = \half B_1 \lambda_1 \ee^{\lambda_1 t} + \half B_2 \lambda_2 \ee^{\lambda_2 t} + \half B_3 \lambda_3 \ee^{\lambda_3 t},
\label{Solution_2:cyv}
\eeqa
\label{Solution:2}
\end{subequations}
where
\beqa
& & F_i = \half \lambda_i^2 + \frac{\Gamma}{2} \lambda_i + \omega^2, \ \ i=1,2,3.
\label{Definition:Fi}
\eeqa
$B_i$, $i=1,2,3$ are constants that can be determined by the initial condition [\eqnref{InitialState:eq}]:
\beq
\left\lbrace \begin{aligned}
& B_1 + B_2 + B_3 = 0,\\
& F_1 B_1 + F_2 B_2 + F_3 B_3 = 0,\\
& B_1 \lambda_1 + B_2 \lambda_2 + B_3 \lambda_3 = 0.
\end{aligned} \right.
\eeq
Thus we obtain
\beq
B_1 = B_2 = B_3 = 0,
\eeq
and
\begin{subequations}
\beqa
& & \sigma^2_y(t|eq) = \frac{1}{\beta m \omega^2},
\label{ASolution_Eq_U:yvar}\\
& & \sigma^2_v(t|eq) = \frac{1}{\beta m},
\label{ASolution_Eq_U:vvar}\\
& & c_{yv}(t|eq) = 0,
\label{ASolution_Eq_U:cyv}
\eeqa
\label{ASolution:2_Eq_U}
\end{subequations}
which is \eqnref{Solution:2_Eq_U}. Equation (\ref{ASolution_Eq_U:yvar}) agrees with the solution to the ESE obtained in Ref.~\cite{imparato2007work}.

For the moments with respect to work, by combining Eqs.~(\ref{EOM:f}), (\ref{EOM:h}), and (\ref{EOM:i}), we obtain a second-order ODE of $\dot{\sigma^2_w}$:
\beq
\dddot{\sigma^2_w} + \Gamma \ddot{\sigma^2_w} + \omega^2 \dot{\sigma^2_w} = 3 m^2 u^2 \omega^4 \dot{\sigma^2_y} + 2 m^2 u^2 \omega^4 \Gamma \sigma^2_y.
\label{Couple:3}
\eeq
By substituting \eqnref{Solution_Eq_U:yvar} into it, we have
\beq
\dddot{\sigma^2_w} + \Gamma \ddot{\sigma^2_w} + \omega^2 \dot{\sigma^2_w} = \frac{2m u^2 \omega^2 \Gamma}{\beta},
\label{Couple:3_Eq}
\eeq
whose characteristic roots are $\mu_{1,2}$ [\eqnref{HO:Mu}]. Equations (\ref{Couple:3_Eq}) and (\ref{EOM:f}), (\ref{EOM:h}), and (\ref{EOM:i}) give the solutions of the moments about $w$,
\begin{subequations}
\beqa
& & \sigma^2_w(t|eq) = \frac{D_1}{\mu_1}\ee^{\mu_1 t} + \frac{D_2}{\mu_2}\ee^{\mu_2 t} + \frac{2m u^2 \Gamma}{\beta}t + E,
\label{Solution_Eq_U:wvar}\\
& & c_{yw}(t|eq) = -\frac{1}{2 m u \omega^2} \rk{D_1 \ee^{\mu_1 t} + D_2 \ee^{\mu_2 t} + \frac{2m u^2 \Gamma}{\beta}},
\label{Solution_Eq_U:cyw}\\
& & c_{vw}(t|eq) = \frac{u}{\beta} - \frac{1}{2 m u \omega^2} \rk{D_1 \mu_1 \ee^{\mu_1 t} + D_2 \mu_2 \ee^{\mu_2 t}},
\label{Solution_Eq_U:cvw}
\eeqa
\label{Solution:3_Eq_U}
\end{subequations}
where $D_1$, $D_2$, $E$ are constants that can be determined by the initial condition [\eqnref{InitialState:eq}]:
\beq
\left\lbrace \begin{aligned}
& \frac{D_1}{\mu_1} + \frac{D_2}{\mu_2} + E = 0,\\
& D_1 + D_2 + \frac{2m u^2 \Gamma}{\beta} = 0,\\
& D_1 \mu_1 + D_2 \mu_2 = \frac{2m u^2 \omega^2}{\beta}.
\end{aligned} \right.
\eeq
Thus we obtain
\beqa
& & D_1 = \frac{2mu \omega^2}{\beta} \rk{\frac{u + \frac{u\Gamma}{\omega^2}\mu_2}{\mu_1 - \mu_2}} = -\frac{2mu \omega^2}{\beta} C_1 \rightarrow -\frac{2m u^2 \Gamma}{\beta}, \nonumber\\
& & D_2 = -\frac{2mu \omega^2}{\beta} \rk{\frac{u + \frac{u\Gamma}{\omega^2}\mu_1}{\mu_1 - \mu_2}} = -\frac{2mu \omega^2}{\beta} C_2 \rightarrow 0, \nonumber\\
& & E = -\frac{D_1}{\mu_1} - \frac{D_2}{\mu_2} \rightarrow -\frac{2m u^2 \Gamma^2}{\beta \omega^2},
\label{D1D2E:Eq}
\eeqa
and
\begin{subequations}
\beqa
& & \sigma^2_w(t|eq) \rightarrow \frac{2m u^2 \Gamma^2}{\beta \omega^2} \rk{\ee^{-\frac{\omega^2}{\Gamma} t}-1 + \frac{\omega^2}{\Gamma}t},
\label{ASolution_Eq_O:wvar}\\
& & c_{yw}(t|eq) \rightarrow \frac{u\Gamma}{\beta \omega^2} \rk{\ee^{-\frac{\omega^2}{\Gamma} t}-1},
\label{ASolution_Eq_O:cyw}
\eeqa
\label{ASolution:3_Eq_O}
\end{subequations}
which is \eqnref{Solution:3_Eq_O}. The r.h.s. of \eqnref{ASolution:3_Eq_O} is identical to the solutions to the ESE obtained in Ref.~\cite{imparato2007work}.

\bibliography{OARefs}

\begin{thebibliography}{57}%
\makeatletter
\providecommand \@ifxundefined [1]{%
 \@ifx{#1\undefined}
}%
\providecommand \@ifnum [1]{%
 \ifnum #1\expandafter \@firstoftwo
 \else \expandafter \@secondoftwo
 \fi
}%
\providecommand \@ifx [1]{%
 \ifx #1\expandafter \@firstoftwo
 \else \expandafter \@secondoftwo
 \fi
}%
\providecommand \natexlab [1]{#1}%
\providecommand \enquote  [1]{``#1''}%
\providecommand \bibnamefont  [1]{#1}%
\providecommand \bibfnamefont [1]{#1}%
\providecommand \citenamefont [1]{#1}%
\providecommand \href@noop [0]{\@secondoftwo}%
\providecommand \href [0]{\begingroup \@sanitize@url \@href}%
\providecommand \@href[1]{\@@startlink{#1}\@@href}%
\providecommand \@@href[1]{\endgroup#1\@@endlink}%
\providecommand \@sanitize@url [0]{\catcode `\\12\catcode `\$12\catcode
  `\&12\catcode `\#12\catcode `\^12\catcode `\_12\catcode `\%12\relax}%
\providecommand \@@startlink[1]{}%
\providecommand \@@endlink[0]{}%
\providecommand \url  [0]{\begingroup\@sanitize@url \@url }%
\providecommand \@url [1]{\endgroup\@href {#1}{\urlprefix }}%
\providecommand \urlprefix  [0]{URL }%
\providecommand \Eprint [0]{\href }%
\providecommand \doibase [0]{http://dx.doi.org/}%
\providecommand \selectlanguage [0]{\@gobble}%
\providecommand \bibinfo  [0]{\@secondoftwo}%
\providecommand \bibfield  [0]{\@secondoftwo}%
\providecommand \translation [1]{[#1]}%
\providecommand \BibitemOpen [0]{}%
\providecommand \bibitemStop [0]{}%
\providecommand \bibitemNoStop [0]{.\EOS\space}%
\providecommand \EOS [0]{\spacefactor3000\relax}%
\providecommand \BibitemShut  [1]{\csname bibitem#1\endcsname}%
\let\auto@bib@innerbib\@empty
\bibitem [{\citenamefont {Purcell}(1977)}]{purcell1977life}%
  \BibitemOpen
  \bibfield  {author} {\bibinfo {author} {\bibfnamefont {E.~M.}\ \bibnamefont
  {Purcell}},\ }\href@noop {} {\bibfield  {journal} {\bibinfo  {journal}
  {American journal of physics}\ }\textbf {\bibinfo {volume} {45}},\ \bibinfo
  {pages} {3} (\bibinfo {year} {1977})}\BibitemShut {NoStop}%
\bibitem [{\citenamefont {Risken}(1996)}]{risken1996fokker}%
  \BibitemOpen
  \bibfield  {author} {\bibinfo {author} {\bibfnamefont {H.}~\bibnamefont
  {Risken}},\ }\href@noop {} {\emph {\bibinfo {title} {The Fokker-Planck
  Equation: Methods of Solution and Applications}}}\ (\bibinfo  {publisher}
  {Springer-Verlag},\ \bibinfo {address} {Berlin},\ \bibinfo {year}
  {1996})\BibitemShut {NoStop}%
\bibitem [{\citenamefont {Kramers}(1940)}]{kramers1940brownian}%
  \BibitemOpen
  \bibfield  {author} {\bibinfo {author} {\bibfnamefont {H.~A.}\ \bibnamefont
  {Kramers}},\ }\href@noop {} {\bibfield  {journal} {\bibinfo  {journal}
  {Physica}\ }\textbf {\bibinfo {volume} {7}},\ \bibinfo {pages} {284}
  (\bibinfo {year} {1940})}\BibitemShut {NoStop}%
\bibitem [{\citenamefont {Von~Smoluchowski}(1916)}]{Smoluchowski1916brownian}%
  \BibitemOpen
  \bibfield  {author} {\bibinfo {author} {\bibfnamefont {M.}~\bibnamefont
  {Von~Smoluchowski}},\ }\href@noop {} {\bibfield  {journal} {\bibinfo
  {journal} {Ann. Physik}\ }\textbf {\bibinfo {volume} {353}},\ \bibinfo
  {pages} {1103} (\bibinfo {year} {1916})}\BibitemShut {NoStop}%
\bibitem [{\citenamefont {Wilemski}(1976)}]{wilemski1976derivation}%
  \BibitemOpen
  \bibfield  {author} {\bibinfo {author} {\bibfnamefont {G.}~\bibnamefont
  {Wilemski}},\ }\href@noop {} {\bibfield  {journal} {\bibinfo  {journal}
  {Journal of Statistical Physics}\ }\textbf {\bibinfo {volume} {14}},\
  \bibinfo {pages} {153} (\bibinfo {year} {1976})}\BibitemShut {NoStop}%
\bibitem [{\citenamefont {Titulaer}(1978)}]{titulaer1978systematic}%
  \BibitemOpen
  \bibfield  {author} {\bibinfo {author} {\bibfnamefont {U.~M.}\ \bibnamefont
  {Titulaer}},\ }\href@noop {} {\bibfield  {journal} {\bibinfo  {journal}
  {Physica A}\ }\textbf {\bibinfo {volume} {91}},\ \bibinfo {pages} {321}
  (\bibinfo {year} {1978})}\BibitemShut {NoStop}%
\bibitem [{\citenamefont {Van~Kampen}(1992)}]{van1992stochastic}%
  \BibitemOpen
  \bibfield  {author} {\bibinfo {author} {\bibfnamefont {N.~G.}\ \bibnamefont
  {Van~Kampen}},\ }\href@noop {} {\emph {\bibinfo {title} {Stochastic Processes
  in Physics and Chemistry}}},\ Vol.~\bibinfo {volume} {1}\ (\bibinfo
  {publisher} {Elsevier},\ \bibinfo {address} {Amsterdam},\ \bibinfo {year}
  {1992})\BibitemShut {NoStop}%
\bibitem [{\citenamefont {Bocquet}(1997)}]{bocquet1997high}%
  \BibitemOpen
  \bibfield  {author} {\bibinfo {author} {\bibfnamefont {L.}~\bibnamefont
  {Bocquet}},\ }\href@noop {} {\bibfield  {journal} {\bibinfo  {journal}
  {American Journal of Physics}\ }\textbf {\bibinfo {volume} {65}},\ \bibinfo
  {pages} {140} (\bibinfo {year} {1997})}\BibitemShut {NoStop}%
\bibitem [{\citenamefont {Jarzynski}(1997)}]{jarzynski1997nonequilibrium}%
  \BibitemOpen
  \bibfield  {author} {\bibinfo {author} {\bibfnamefont {C.}~\bibnamefont
  {Jarzynski}},\ }\href@noop {} {\bibfield  {journal} {\bibinfo  {journal}
  {Physical Review Letters}\ }\textbf {\bibinfo {volume} {78}},\ \bibinfo
  {pages} {2690} (\bibinfo {year} {1997})}\BibitemShut {NoStop}%
\bibitem [{\citenamefont {Crooks}(1999)}]{crooks1999entropy}%
  \BibitemOpen
  \bibfield  {author} {\bibinfo {author} {\bibfnamefont {G.~E.}\ \bibnamefont
  {Crooks}},\ }\href@noop {} {\bibfield  {journal} {\bibinfo  {journal}
  {Physical Review E}\ }\textbf {\bibinfo {volume} {60}},\ \bibinfo {pages}
  {2721} (\bibinfo {year} {1999})}\BibitemShut {NoStop}%
\bibitem [{\citenamefont {Sekimoto}(1998)}]{sekimoto1998langevin}%
  \BibitemOpen
  \bibfield  {author} {\bibinfo {author} {\bibfnamefont {K.}~\bibnamefont
  {Sekimoto}},\ }\href@noop {} {\bibfield  {journal} {\bibinfo  {journal}
  {Progress of Theoretical Physics Supplement}\ }\textbf {\bibinfo {volume}
  {130}},\ \bibinfo {pages} {17} (\bibinfo {year} {1998})}\BibitemShut
  {NoStop}%
\bibitem [{\citenamefont {Sekimoto}(2010)}]{sekimoto2010stochastic}%
  \BibitemOpen
  \bibfield  {author} {\bibinfo {author} {\bibfnamefont {K.}~\bibnamefont
  {Sekimoto}},\ }\href@noop {} {\emph {\bibinfo {title} {Stochastic
  Energetics}}},\ \bibinfo {series} {Lecture Notes in Physics}, Vol.\ \bibinfo
  {volume} {799}\ (\bibinfo  {publisher} {Springer-Verlag},\ \bibinfo {address}
  {Berlin},\ \bibinfo {year} {2010})\BibitemShut {NoStop}%
\bibitem [{\citenamefont {Seifert}(2005)}]{seifert2005entropy}%
  \BibitemOpen
  \bibfield  {author} {\bibinfo {author} {\bibfnamefont {U.}~\bibnamefont
  {Seifert}},\ }\href@noop {} {\bibfield  {journal} {\bibinfo  {journal}
  {Physical review letters}\ }\textbf {\bibinfo {volume} {95}},\ \bibinfo
  {pages} {040602} (\bibinfo {year} {2005})}\BibitemShut {NoStop}%
\bibitem [{\citenamefont {Seifert}(2012)}]{seifert2012stochastic}%
  \BibitemOpen
  \bibfield  {author} {\bibinfo {author} {\bibfnamefont {U.}~\bibnamefont
  {Seifert}},\ }\href@noop {} {\bibfield  {journal} {\bibinfo  {journal}
  {Reports on Progress in Physics}\ }\textbf {\bibinfo {volume} {75}},\
  \bibinfo {pages} {126001} (\bibinfo {year} {2012})}\BibitemShut {NoStop}%
\bibitem [{\citenamefont {Liphardt}\ \emph {et~al.}(2002)\citenamefont
  {Liphardt}, \citenamefont {Dumont}, \citenamefont {Smith}, \citenamefont
  {Tinoco},\ and\ \citenamefont {Bustamante}}]{liphardt2002equilibrium}%
  \BibitemOpen
  \bibfield  {author} {\bibinfo {author} {\bibfnamefont {J.}~\bibnamefont
  {Liphardt}}, \bibinfo {author} {\bibfnamefont {S.}~\bibnamefont {Dumont}},
  \bibinfo {author} {\bibfnamefont {S.~B.}\ \bibnamefont {Smith}}, \bibinfo
  {author} {\bibfnamefont {I.}~\bibnamefont {Tinoco}}, \ and\ \bibinfo {author}
  {\bibfnamefont {C.}~\bibnamefont {Bustamante}},\ }\href@noop {} {\bibfield
  {journal} {\bibinfo  {journal} {Science}\ }\textbf {\bibinfo {volume}
  {296}},\ \bibinfo {pages} {1832} (\bibinfo {year} {2002})}\BibitemShut
  {NoStop}%
\bibitem [{\citenamefont {Collin}\ \emph {et~al.}(2005)\citenamefont {Collin},
  \citenamefont {Ritort}, \citenamefont {Jarzynski}, \citenamefont {Smith},
  \citenamefont {Tinoco},\ and\ \citenamefont
  {Bustamante}}]{collin2005verification}%
  \BibitemOpen
  \bibfield  {author} {\bibinfo {author} {\bibfnamefont {D.}~\bibnamefont
  {Collin}}, \bibinfo {author} {\bibfnamefont {F.}~\bibnamefont {Ritort}},
  \bibinfo {author} {\bibfnamefont {C.}~\bibnamefont {Jarzynski}}, \bibinfo
  {author} {\bibfnamefont {S.~B.}\ \bibnamefont {Smith}}, \bibinfo {author}
  {\bibfnamefont {I.}~\bibnamefont {Tinoco}}, \ and\ \bibinfo {author}
  {\bibfnamefont {C.}~\bibnamefont {Bustamante}},\ }\href@noop {} {\bibfield
  {journal} {\bibinfo  {journal} {Nature (London)}\ }\textbf {\bibinfo {volume}
  {437}},\ \bibinfo {pages} {231} (\bibinfo {year} {2005})}\BibitemShut
  {NoStop}%
\bibitem [{\citenamefont {Douarche}\ \emph {et~al.}(2006)\citenamefont
  {Douarche}, \citenamefont {Joubaud}, \citenamefont {Garnier}, \citenamefont
  {Petrosyan},\ and\ \citenamefont {Ciliberto}}]{douarche2006work}%
  \BibitemOpen
  \bibfield  {author} {\bibinfo {author} {\bibfnamefont {F.}~\bibnamefont
  {Douarche}}, \bibinfo {author} {\bibfnamefont {S.}~\bibnamefont {Joubaud}},
  \bibinfo {author} {\bibfnamefont {N.~B.}\ \bibnamefont {Garnier}}, \bibinfo
  {author} {\bibfnamefont {A.}~\bibnamefont {Petrosyan}}, \ and\ \bibinfo
  {author} {\bibfnamefont {S.}~\bibnamefont {Ciliberto}},\ }\href@noop {}
  {\bibfield  {journal} {\bibinfo  {journal} {Physical review letters}\
  }\textbf {\bibinfo {volume} {97}},\ \bibinfo {pages} {140603} (\bibinfo
  {year} {2006})}\BibitemShut {NoStop}%
\bibitem [{\citenamefont {Junier}\ \emph {et~al.}(2009)\citenamefont {Junier},
  \citenamefont {Mossa}, \citenamefont {Manosas},\ and\ \citenamefont
  {Ritort}}]{junier2009recovery}%
  \BibitemOpen
  \bibfield  {author} {\bibinfo {author} {\bibfnamefont {I.}~\bibnamefont
  {Junier}}, \bibinfo {author} {\bibfnamefont {A.}~\bibnamefont {Mossa}},
  \bibinfo {author} {\bibfnamefont {M.}~\bibnamefont {Manosas}}, \ and\
  \bibinfo {author} {\bibfnamefont {F.}~\bibnamefont {Ritort}},\ }\href@noop {}
  {\bibfield  {journal} {\bibinfo  {journal} {Physical review letters}\
  }\textbf {\bibinfo {volume} {102}},\ \bibinfo {pages} {070602} (\bibinfo
  {year} {2009})}\BibitemShut {NoStop}%
\bibitem [{\citenamefont {Mart{\'\i}nez}\ \emph {et~al.}(2016)\citenamefont
  {Mart{\'\i}nez}, \citenamefont {Rold{\'a}n}, \citenamefont {Dinis},
  \citenamefont {Petrov}, \citenamefont {Parrondo},\ and\ \citenamefont
  {Rica}}]{martinez2016brownian}%
  \BibitemOpen
  \bibfield  {author} {\bibinfo {author} {\bibfnamefont {I.~A.}\ \bibnamefont
  {Mart{\'\i}nez}}, \bibinfo {author} {\bibfnamefont {{\'E}.}~\bibnamefont
  {Rold{\'a}n}}, \bibinfo {author} {\bibfnamefont {L.}~\bibnamefont {Dinis}},
  \bibinfo {author} {\bibfnamefont {D.}~\bibnamefont {Petrov}}, \bibinfo
  {author} {\bibfnamefont {J.~M.~R.}\ \bibnamefont {Parrondo}}, \ and\ \bibinfo
  {author} {\bibfnamefont {R.~A.}\ \bibnamefont {Rica}},\ }\href@noop {}
  {\bibfield  {journal} {\bibinfo  {journal} {Nature physics}\ }\textbf
  {\bibinfo {volume} {12}},\ \bibinfo {pages} {67} (\bibinfo {year}
  {2016})}\BibitemShut {NoStop}%
\bibitem [{\citenamefont {Toyabe}\ \emph {et~al.}(2010)\citenamefont {Toyabe},
  \citenamefont {Sagawa}, \citenamefont {Ueda}, \citenamefont {Muneyuki},\ and\
  \citenamefont {Sano}}]{toyabe2010experimental}%
  \BibitemOpen
  \bibfield  {author} {\bibinfo {author} {\bibfnamefont {S.}~\bibnamefont
  {Toyabe}}, \bibinfo {author} {\bibfnamefont {T.}~\bibnamefont {Sagawa}},
  \bibinfo {author} {\bibfnamefont {M.}~\bibnamefont {Ueda}}, \bibinfo {author}
  {\bibfnamefont {E.}~\bibnamefont {Muneyuki}}, \ and\ \bibinfo {author}
  {\bibfnamefont {M.}~\bibnamefont {Sano}},\ }\href@noop {} {\bibfield
  {journal} {\bibinfo  {journal} {Nature physics}\ }\textbf {\bibinfo {volume}
  {6}},\ \bibinfo {pages} {988} (\bibinfo {year} {2010})}\BibitemShut {NoStop}%
\bibitem [{\citenamefont {Celani}\ \emph {et~al.}(2012)\citenamefont {Celani},
  \citenamefont {Bo}, \citenamefont {Eichhorn},\ and\ \citenamefont
  {Aurell}}]{celani2012anomalous}%
  \BibitemOpen
  \bibfield  {author} {\bibinfo {author} {\bibfnamefont {A.}~\bibnamefont
  {Celani}}, \bibinfo {author} {\bibfnamefont {S.}~\bibnamefont {Bo}}, \bibinfo
  {author} {\bibfnamefont {R.}~\bibnamefont {Eichhorn}}, \ and\ \bibinfo
  {author} {\bibfnamefont {E.}~\bibnamefont {Aurell}},\ }\href@noop {}
  {\bibfield  {journal} {\bibinfo  {journal} {Physical review letters}\
  }\textbf {\bibinfo {volume} {109}},\ \bibinfo {pages} {260603} (\bibinfo
  {year} {2012})}\BibitemShut {NoStop}%
\bibitem [{\citenamefont {Bo}\ and\ \citenamefont
  {Celani}(2013)}]{bo2013entropic}%
  \BibitemOpen
  \bibfield  {author} {\bibinfo {author} {\bibfnamefont {S.}~\bibnamefont
  {Bo}}\ and\ \bibinfo {author} {\bibfnamefont {A.}~\bibnamefont {Celani}},\
  }\href@noop {} {\bibfield  {journal} {\bibinfo  {journal} {Physical Review
  E}\ }\textbf {\bibinfo {volume} {87}},\ \bibinfo {pages} {050102} (\bibinfo
  {year} {2013})}\BibitemShut {NoStop}%
\bibitem [{\citenamefont {Bo}\ and\ \citenamefont
  {Celani}(2017)}]{bo2016multiple}%
  \BibitemOpen
  \bibfield  {author} {\bibinfo {author} {\bibfnamefont {S.}~\bibnamefont
  {Bo}}\ and\ \bibinfo {author} {\bibfnamefont {A.}~\bibnamefont {Celani}},\
  }\href@noop {} {\bibfield  {journal} {\bibinfo  {journal} {Physics reports}\
  }\textbf {\bibinfo {volume} {670}},\ \bibinfo {pages} {1} (\bibinfo {year}
  {2017})}\BibitemShut {NoStop}%
\bibitem [{\citenamefont {Ge}(2014)}]{ge2014time}%
  \BibitemOpen
  \bibfield  {author} {\bibinfo {author} {\bibfnamefont {H.}~\bibnamefont
  {Ge}},\ }\href@noop {} {\bibfield  {journal} {\bibinfo  {journal} {Physical
  Review E}\ }\textbf {\bibinfo {volume} {89}},\ \bibinfo {pages} {022127}
  (\bibinfo {year} {2014})}\BibitemShut {NoStop}%
\bibitem [{\citenamefont {Mart{\'\i}nez}\ \emph {et~al.}(2015)\citenamefont
  {Mart{\'\i}nez}, \citenamefont {Rold{\'a}n}, \citenamefont {Dinis},
  \citenamefont {Petrov},\ and\ \citenamefont {Rica}}]{martinez2015adiabatic}%
  \BibitemOpen
  \bibfield  {author} {\bibinfo {author} {\bibfnamefont {I.~A.}\ \bibnamefont
  {Mart{\'\i}nez}}, \bibinfo {author} {\bibfnamefont {{\'E}.}~\bibnamefont
  {Rold{\'a}n}}, \bibinfo {author} {\bibfnamefont {L.}~\bibnamefont {Dinis}},
  \bibinfo {author} {\bibfnamefont {D.}~\bibnamefont {Petrov}}, \ and\ \bibinfo
  {author} {\bibfnamefont {R.~A.}\ \bibnamefont {Rica}},\ }\href@noop {}
  {\bibfield  {journal} {\bibinfo  {journal} {Physical review letters}\
  }\textbf {\bibinfo {volume} {114}},\ \bibinfo {pages} {120601} (\bibinfo
  {year} {2015})}\BibitemShut {NoStop}%
\bibitem [{\citenamefont {Kawaguchi}\ and\ \citenamefont
  {Nakayama}(2013)}]{kawaguchi2013fluctuation}%
  \BibitemOpen
  \bibfield  {author} {\bibinfo {author} {\bibfnamefont {K.}~\bibnamefont
  {Kawaguchi}}\ and\ \bibinfo {author} {\bibfnamefont {Y.}~\bibnamefont
  {Nakayama}},\ }\href@noop {} {\bibfield  {journal} {\bibinfo  {journal}
  {Physical Review E}\ }\textbf {\bibinfo {volume} {88}},\ \bibinfo {pages}
  {022147} (\bibinfo {year} {2013})}\BibitemShut {NoStop}%
\bibitem [{\citenamefont {Lan}\ and\ \citenamefont
  {Aurell}(2015)}]{lan2015stochastic}%
  \BibitemOpen
  \bibfield  {author} {\bibinfo {author} {\bibfnamefont {Y.}~\bibnamefont
  {Lan}}\ and\ \bibinfo {author} {\bibfnamefont {E.}~\bibnamefont {Aurell}},\
  }\href@noop {} {\bibfield  {journal} {\bibinfo  {journal} {Scientific
  reports}\ }\textbf {\bibinfo {volume} {5}},\ \bibinfo {pages} {12266}
  (\bibinfo {year} {2015})}\BibitemShut {NoStop}%
\bibitem [{\citenamefont {Marino}\ \emph {et~al.}(2016)\citenamefont {Marino},
  \citenamefont {Eichhorn},\ and\ \citenamefont {Aurell}}]{marino2016entropy}%
  \BibitemOpen
  \bibfield  {author} {\bibinfo {author} {\bibfnamefont {R.}~\bibnamefont
  {Marino}}, \bibinfo {author} {\bibfnamefont {R.}~\bibnamefont {Eichhorn}}, \
  and\ \bibinfo {author} {\bibfnamefont {E.}~\bibnamefont {Aurell}},\
  }\href@noop {} {\bibfield  {journal} {\bibinfo  {journal} {Physical Review
  E}\ }\textbf {\bibinfo {volume} {93}},\ \bibinfo {pages} {012132} (\bibinfo
  {year} {2016})}\BibitemShut {NoStop}%
\bibitem [{\citenamefont {Taniguchi}\ and\ \citenamefont
  {Cohen}(2008)}]{taniguchi2008inertial}%
  \BibitemOpen
  \bibfield  {author} {\bibinfo {author} {\bibfnamefont {T.}~\bibnamefont
  {Taniguchi}}\ and\ \bibinfo {author} {\bibfnamefont {E.~G.~D.}\ \bibnamefont
  {Cohen}},\ }\href@noop {} {\bibfield  {journal} {\bibinfo  {journal} {Journal
  of Statistical Physics}\ }\textbf {\bibinfo {volume} {130}},\ \bibinfo
  {pages} {1} (\bibinfo {year} {2008})}\BibitemShut {NoStop}%
\bibitem [{\citenamefont {Kwon}\ \emph {et~al.}(2013)\citenamefont {Kwon},
  \citenamefont {Noh},\ and\ \citenamefont {Park}}]{kwon2013work}%
  \BibitemOpen
  \bibfield  {author} {\bibinfo {author} {\bibfnamefont {C.}~\bibnamefont
  {Kwon}}, \bibinfo {author} {\bibfnamefont {J.~D.}\ \bibnamefont {Noh}}, \
  and\ \bibinfo {author} {\bibfnamefont {H.}~\bibnamefont {Park}},\ }\href@noop
  {} {\bibfield  {journal} {\bibinfo  {journal} {Physical Review E}\ }\textbf
  {\bibinfo {volume} {88}},\ \bibinfo {pages} {062102} (\bibinfo {year}
  {2013})}\BibitemShut {NoStop}%
\bibitem [{\citenamefont {Imparato}\ and\ \citenamefont
  {Peliti}(2006)}]{imparato2006fluctuation}%
  \BibitemOpen
  \bibfield  {author} {\bibinfo {author} {\bibfnamefont {A.}~\bibnamefont
  {Imparato}}\ and\ \bibinfo {author} {\bibfnamefont {L.}~\bibnamefont
  {Peliti}},\ }\href@noop {} {\bibfield  {journal} {\bibinfo  {journal}
  {Physical Review E}\ }\textbf {\bibinfo {volume} {74}},\ \bibinfo {pages}
  {026106} (\bibinfo {year} {2006})}\BibitemShut {NoStop}%
\bibitem [{\citenamefont {Gomez-Marin}\ \emph {et~al.}(2008)\citenamefont
  {Gomez-Marin}, \citenamefont {Schmiedl},\ and\ \citenamefont
  {Seifert}}]{gomez2008optimal}%
  \BibitemOpen
  \bibfield  {author} {\bibinfo {author} {\bibfnamefont {A.}~\bibnamefont
  {Gomez-Marin}}, \bibinfo {author} {\bibfnamefont {T.}~\bibnamefont
  {Schmiedl}}, \ and\ \bibinfo {author} {\bibfnamefont {U.}~\bibnamefont
  {Seifert}},\ }\href@noop {} {\bibfield  {journal} {\bibinfo  {journal} {The
  Journal of chemical physics}\ }\textbf {\bibinfo {volume} {129}},\ \bibinfo
  {pages} {024114} (\bibinfo {year} {2008})}\BibitemShut {NoStop}%
\bibitem [{\citenamefont {Baule}\ and\ \citenamefont
  {Cohen}(2009)}]{baule2009steady}%
  \BibitemOpen
  \bibfield  {author} {\bibinfo {author} {\bibfnamefont {A.}~\bibnamefont
  {Baule}}\ and\ \bibinfo {author} {\bibfnamefont {E.~G.~D.}\ \bibnamefont
  {Cohen}},\ }\href@noop {} {\bibfield  {journal} {\bibinfo  {journal}
  {Physical Review E}\ }\textbf {\bibinfo {volume} {80}},\ \bibinfo {pages}
  {011110} (\bibinfo {year} {2009})}\BibitemShut {NoStop}%
\bibitem [{\citenamefont {Minh}\ and\ \citenamefont
  {Adib}(2009)}]{minh2009path}%
  \BibitemOpen
  \bibfield  {author} {\bibinfo {author} {\bibfnamefont {D.~D.~L.}\
  \bibnamefont {Minh}}\ and\ \bibinfo {author} {\bibfnamefont {A.~B.}\
  \bibnamefont {Adib}},\ }\href@noop {} {\bibfield  {journal} {\bibinfo
  {journal} {Physical Review E}\ }\textbf {\bibinfo {volume} {79}},\ \bibinfo
  {pages} {021122} (\bibinfo {year} {2009})}\BibitemShut {NoStop}%
\bibitem [{\citenamefont {Engel}(2009)}]{engel2009asymptotics}%
  \BibitemOpen
  \bibfield  {author} {\bibinfo {author} {\bibfnamefont {A.}~\bibnamefont
  {Engel}},\ }\href@noop {} {\bibfield  {journal} {\bibinfo  {journal}
  {Physical Review E}\ }\textbf {\bibinfo {volume} {80}},\ \bibinfo {pages}
  {021120} (\bibinfo {year} {2009})}\BibitemShut {NoStop}%
\bibitem [{\citenamefont {Holubec}\ \emph {et~al.}(2015)\citenamefont
  {Holubec}, \citenamefont {Dierl}, \citenamefont {Einax}, \citenamefont
  {Maass}, \citenamefont {Chvosta},\ and\ \citenamefont
  {Ryabov}}]{holubec2015asymptotics}%
  \BibitemOpen
  \bibfield  {author} {\bibinfo {author} {\bibfnamefont {V.}~\bibnamefont
  {Holubec}}, \bibinfo {author} {\bibfnamefont {M.}~\bibnamefont {Dierl}},
  \bibinfo {author} {\bibfnamefont {M.}~\bibnamefont {Einax}}, \bibinfo
  {author} {\bibfnamefont {P.}~\bibnamefont {Maass}}, \bibinfo {author}
  {\bibfnamefont {P.}~\bibnamefont {Chvosta}}, \ and\ \bibinfo {author}
  {\bibfnamefont {A.}~\bibnamefont {Ryabov}},\ }\href@noop {} {\bibfield
  {journal} {\bibinfo  {journal} {Physica Scripta}\ }\textbf {\bibinfo {volume}
  {2015}},\ \bibinfo {pages} {014024} (\bibinfo {year} {2015})}\BibitemShut
  {NoStop}%
\bibitem [{\citenamefont {Li}\ and\ \citenamefont
  {Raizen}(2013)}]{li2013brownian}%
  \BibitemOpen
  \bibfield  {author} {\bibinfo {author} {\bibfnamefont {T.}~\bibnamefont
  {Li}}\ and\ \bibinfo {author} {\bibfnamefont {M.~G.}\ \bibnamefont
  {Raizen}},\ }\href@noop {} {\bibfield  {journal} {\bibinfo  {journal}
  {Annalen der Physik}\ }\textbf {\bibinfo {volume} {525}},\ \bibinfo {pages}
  {281} (\bibinfo {year} {2013})}\BibitemShut {NoStop}%
\bibitem [{\citenamefont {Esposito}\ and\ \citenamefont
  {Parrondo}(2015)}]{esposito2015stochastic}%
  \BibitemOpen
  \bibfield  {author} {\bibinfo {author} {\bibfnamefont {M.}~\bibnamefont
  {Esposito}}\ and\ \bibinfo {author} {\bibfnamefont {J.~M.~R.}\ \bibnamefont
  {Parrondo}},\ }\href@noop {} {\bibfield  {journal} {\bibinfo  {journal}
  {Physical Review E}\ }\textbf {\bibinfo {volume} {91}},\ \bibinfo {pages}
  {052114} (\bibinfo {year} {2015})}\BibitemShut {NoStop}%
\bibitem [{\citenamefont {Spinney}\ and\ \citenamefont
  {Ford}(2012)}]{spinney2012entropy}%
  \BibitemOpen
  \bibfield  {author} {\bibinfo {author} {\bibfnamefont {R.~E.}\ \bibnamefont
  {Spinney}}\ and\ \bibinfo {author} {\bibfnamefont {I.~J.}\ \bibnamefont
  {Ford}},\ }\href@noop {} {\bibfield  {journal} {\bibinfo  {journal} {Physical
  Review E}\ }\textbf {\bibinfo {volume} {85}},\ \bibinfo {pages} {051113}
  (\bibinfo {year} {2012})}\BibitemShut {NoStop}%
\bibitem [{\citenamefont {Kim}\ and\ \citenamefont
  {Qian}(2004)}]{kim2004entropy}%
  \BibitemOpen
  \bibfield  {author} {\bibinfo {author} {\bibfnamefont {K.~H.}\ \bibnamefont
  {Kim}}\ and\ \bibinfo {author} {\bibfnamefont {H.}~\bibnamefont {Qian}},\
  }\href@noop {} {\bibfield  {journal} {\bibinfo  {journal} {Physical review
  letters}\ }\textbf {\bibinfo {volume} {93}},\ \bibinfo {pages} {120602}
  (\bibinfo {year} {2004})}\BibitemShut {NoStop}%
\bibitem [{\citenamefont {Zhao}\ and\ \citenamefont
  {Zhao}(2017)}]{zhao2017brownian}%
  \BibitemOpen
  \bibfield  {author} {\bibinfo {author} {\bibfnamefont {H.}~\bibnamefont
  {Zhao}}\ and\ \bibinfo {author} {\bibfnamefont {H.}~\bibnamefont {Zhao}},\
  }\href@noop {} {\bibfield  {journal} {\bibinfo  {journal} {arXiv:1706.00779}\
  } (\bibinfo {year} {2017})}\BibitemShut {NoStop}%
\bibitem [{\citenamefont {Esposito}(2012)}]{esposito2012stochastic}%
  \BibitemOpen
  \bibfield  {author} {\bibinfo {author} {\bibfnamefont {M.}~\bibnamefont
  {Esposito}},\ }\href@noop {} {\bibfield  {journal} {\bibinfo  {journal}
  {Physical Review E}\ }\textbf {\bibinfo {volume} {85}},\ \bibinfo {pages}
  {041125} (\bibinfo {year} {2012})}\BibitemShut {NoStop}%
\bibitem [{\citenamefont {Li}\ \emph {et~al.}(2017)\citenamefont {Li},
  \citenamefont {Quan},\ and\ \citenamefont {Tu}}]{li2017shortcuts}%
  \BibitemOpen
  \bibfield  {author} {\bibinfo {author} {\bibfnamefont {G.}~\bibnamefont
  {Li}}, \bibinfo {author} {\bibfnamefont {H.~T.}\ \bibnamefont {Quan}}, \ and\
  \bibinfo {author} {\bibfnamefont {Z.~C.}\ \bibnamefont {Tu}},\ }\href@noop {}
  {\bibfield  {journal} {\bibinfo  {journal} {Physical Review E}\ }\textbf
  {\bibinfo {volume} {96}},\ \bibinfo {pages} {012144} (\bibinfo {year}
  {2017})}\BibitemShut {NoStop}%
\bibitem [{\citenamefont {Yin}\ and\ \citenamefont
  {Ao}(2006)}]{yin2006existence}%
  \BibitemOpen
  \bibfield  {author} {\bibinfo {author} {\bibfnamefont {L.}~\bibnamefont
  {Yin}}\ and\ \bibinfo {author} {\bibfnamefont {P.}~\bibnamefont {Ao}},\
  }\href@noop {} {\bibfield  {journal} {\bibinfo  {journal} {Journal of Physics
  A: Mathematical and General}\ }\textbf {\bibinfo {volume} {39}},\ \bibinfo
  {pages} {8593} (\bibinfo {year} {2006})}\BibitemShut {NoStop}%
\bibitem [{\citenamefont {Li}\ \emph {et~al.}(2010)\citenamefont {Li},
  \citenamefont {Kheifets}, \citenamefont {Medellin},\ and\ \citenamefont
  {Raizen}}]{li2010measurement}%
  \BibitemOpen
  \bibfield  {author} {\bibinfo {author} {\bibfnamefont {T.}~\bibnamefont
  {Li}}, \bibinfo {author} {\bibfnamefont {S.}~\bibnamefont {Kheifets}},
  \bibinfo {author} {\bibfnamefont {D.}~\bibnamefont {Medellin}}, \ and\
  \bibinfo {author} {\bibfnamefont {M.~G.}\ \bibnamefont {Raizen}},\
  }\href@noop {} {\bibfield  {journal} {\bibinfo  {journal} {Science}\ }\textbf
  {\bibinfo {volume} {328}},\ \bibinfo {pages} {1673} (\bibinfo {year}
  {2010})}\BibitemShut {NoStop}%
\bibitem [{\citenamefont {Kheifets}\ \emph {et~al.}(2014)\citenamefont
  {Kheifets}, \citenamefont {Simha}, \citenamefont {Melin}, \citenamefont
  {Li},\ and\ \citenamefont {Raizen}}]{kheifets2014observation}%
  \BibitemOpen
  \bibfield  {author} {\bibinfo {author} {\bibfnamefont {S.}~\bibnamefont
  {Kheifets}}, \bibinfo {author} {\bibfnamefont {A.}~\bibnamefont {Simha}},
  \bibinfo {author} {\bibfnamefont {K.}~\bibnamefont {Melin}}, \bibinfo
  {author} {\bibfnamefont {T.}~\bibnamefont {Li}}, \ and\ \bibinfo {author}
  {\bibfnamefont {M.~G.}\ \bibnamefont {Raizen}},\ }\href@noop {} {\bibfield
  {journal} {\bibinfo  {journal} {Science}\ }\textbf {\bibinfo {volume}
  {343}},\ \bibinfo {pages} {1493} (\bibinfo {year} {2014})}\BibitemShut
  {NoStop}%
\bibitem [{\citenamefont {Wang}\ \emph {et~al.}(2016)\citenamefont {Wang},
  \citenamefont {Kawaguchi}, \citenamefont {Sasa},\ and\ \citenamefont
  {Tang}}]{wang2016entropy}%
  \BibitemOpen
  \bibfield  {author} {\bibinfo {author} {\bibfnamefont {S.-W.}\ \bibnamefont
  {Wang}}, \bibinfo {author} {\bibfnamefont {K.}~\bibnamefont {Kawaguchi}},
  \bibinfo {author} {\bibfnamefont {S.-i.}\ \bibnamefont {Sasa}}, \ and\
  \bibinfo {author} {\bibfnamefont {L.-H.}\ \bibnamefont {Tang}},\ }\href@noop
  {} {\bibfield  {journal} {\bibinfo  {journal} {Physical review letters}\
  }\textbf {\bibinfo {volume} {117}},\ \bibinfo {pages} {070601} (\bibinfo
  {year} {2016})}\BibitemShut {NoStop}%
\bibitem [{\citenamefont {Mazonka}\ and\ \citenamefont
  {Jarzynski}(1999)}]{mazonka1999exactly}%
  \BibitemOpen
  \bibfield  {author} {\bibinfo {author} {\bibfnamefont {O.}~\bibnamefont
  {Mazonka}}\ and\ \bibinfo {author} {\bibfnamefont {C.}~\bibnamefont
  {Jarzynski}},\ }\href@noop {} {\bibfield  {journal} {\bibinfo  {journal}
  {arXiv:cond-mat/9912121}\ } (\bibinfo {year} {1999})}\BibitemShut {NoStop}%
\bibitem [{\citenamefont {Imparato}\ \emph {et~al.}(2007)\citenamefont
  {Imparato}, \citenamefont {Peliti}, \citenamefont {Pesce}, \citenamefont
  {Rusciano},\ and\ \citenamefont {Sasso}}]{imparato2007work}%
  \BibitemOpen
  \bibfield  {author} {\bibinfo {author} {\bibfnamefont {A.}~\bibnamefont
  {Imparato}}, \bibinfo {author} {\bibfnamefont {L.}~\bibnamefont {Peliti}},
  \bibinfo {author} {\bibfnamefont {G.}~\bibnamefont {Pesce}}, \bibinfo
  {author} {\bibfnamefont {G.}~\bibnamefont {Rusciano}}, \ and\ \bibinfo
  {author} {\bibfnamefont {A.}~\bibnamefont {Sasso}},\ }\href@noop {}
  {\bibfield  {journal} {\bibinfo  {journal} {Physical Review E}\ }\textbf
  {\bibinfo {volume} {76}},\ \bibinfo {pages} {050101} (\bibinfo {year}
  {2007})}\BibitemShut {NoStop}%
\bibitem [{\citenamefont {Taniguchi}\ and\ \citenamefont
  {Cohen}(2007)}]{taniguchi2007onsager}%
  \BibitemOpen
  \bibfield  {author} {\bibinfo {author} {\bibfnamefont {T.}~\bibnamefont
  {Taniguchi}}\ and\ \bibinfo {author} {\bibfnamefont {E.~G.~D.}\ \bibnamefont
  {Cohen}},\ }\href@noop {} {\bibfield  {journal} {\bibinfo  {journal} {Journal
  of Statistical Physics}\ }\textbf {\bibinfo {volume} {126}},\ \bibinfo
  {pages} {1} (\bibinfo {year} {2007})}\BibitemShut {NoStop}%
\bibitem [{\citenamefont {Speck}\ and\ \citenamefont
  {Seifert}(2005)}]{speck2005dissipated}%
  \BibitemOpen
  \bibfield  {author} {\bibinfo {author} {\bibfnamefont {T.}~\bibnamefont
  {Speck}}\ and\ \bibinfo {author} {\bibfnamefont {U.}~\bibnamefont
  {Seifert}},\ }\href@noop {} {\bibfield  {journal} {\bibinfo  {journal} {Eur.
  Phys. J. B}\ }\textbf {\bibinfo {volume} {43}},\ \bibinfo {pages} {521}
  (\bibinfo {year} {2005})}\BibitemShut {NoStop}%
\bibitem [{\citenamefont {Imparato}\ and\ \citenamefont
  {Peliti}(2005)}]{imparato2005work}%
  \BibitemOpen
  \bibfield  {author} {\bibinfo {author} {\bibfnamefont {A.}~\bibnamefont
  {Imparato}}\ and\ \bibinfo {author} {\bibfnamefont {L.}~\bibnamefont
  {Peliti}},\ }\href@noop {} {\bibfield  {journal} {\bibinfo  {journal}
  {Physical Review E}\ }\textbf {\bibinfo {volume} {72}},\ \bibinfo {pages}
  {046114} (\bibinfo {year} {2005})}\BibitemShut {NoStop}%
\bibitem [{arr()}]{arrow}%
  \BibitemOpen
  \href@noop {} {}\bibinfo {note} {Throughout the paper, we use the arrow
  ``$\rightarrow$'' to indicate the asymptotic expressions in the overdamped
  limit.}\BibitemShut {Stop}%
\bibitem [{def()}]{definition}%
  \BibitemOpen
  \href@noop {} {}\bibinfo {note} {$\sigma^2_y = \overline{y^2} - \bar{y}^2$,
  $\sigma^2_v = \overline{v^2} - \bar{v}^2$, $\sigma^2_w = \overline{w^2} -
  \bar{w}^2$, $c_{yv} = \overline{yv} - \bar{y}\bar{v}$, $c_{yw} =
  \overline{yw} - \bar{y}\bar{w}$, $c_{vw} = \overline{vw} - \bar{v}\bar{w}$,
  where the bar above the variable denotes the mean value averaged over
  $p^G(y,v,w,t)$.}\BibitemShut {Stop}%
\bibitem [{\citenamefont {Funo}\ and\ \citenamefont
  {Quan}(2018)}]{funo2018path}%
  \BibitemOpen
  \bibfield  {author} {\bibinfo {author} {\bibfnamefont {K.}~\bibnamefont
  {Funo}}\ and\ \bibinfo {author} {\bibfnamefont {H.~T.}\ \bibnamefont
  {Quan}},\ }\href@noop {} {\bibfield  {journal} {\bibinfo  {journal} {Physical
  Review Letters}\ }\textbf {\bibinfo {volume} {121}},\ \bibinfo {pages}
  {040602} (\bibinfo {year} {2018})}\BibitemShut {NoStop}%
\bibitem [{\citenamefont {Hoang}\ \emph {et~al.}(2018)\citenamefont {Hoang},
  \citenamefont {Pan}, \citenamefont {Ahn}, \citenamefont {Bang}, \citenamefont
  {Quan},\ and\ \citenamefont {Li}}]{hoang2018experimental}%
  \BibitemOpen
  \bibfield  {author} {\bibinfo {author} {\bibfnamefont {T.~M.}\ \bibnamefont
  {Hoang}}, \bibinfo {author} {\bibfnamefont {R.}~\bibnamefont {Pan}}, \bibinfo
  {author} {\bibfnamefont {J.}~\bibnamefont {Ahn}}, \bibinfo {author}
  {\bibfnamefont {J.}~\bibnamefont {Bang}}, \bibinfo {author} {\bibfnamefont
  {H.~T.}\ \bibnamefont {Quan}}, \ and\ \bibinfo {author} {\bibfnamefont
  {T.}~\bibnamefont {Li}},\ }\href@noop {} {\bibfield  {journal} {\bibinfo
  {journal} {Physical Review Letters}\ }\textbf {\bibinfo {volume} {120}},\
  \bibinfo {pages} {080602} (\bibinfo {year} {2018})}\BibitemShut {NoStop}%
\bibitem [{\citenamefont {Hoang}\ \emph {et~al.}(2016)\citenamefont {Hoang},
  \citenamefont {Ahn}, \citenamefont {Bang},\ and\ \citenamefont
  {Li}}]{hoang2016electron}%
  \BibitemOpen
  \bibfield  {author} {\bibinfo {author} {\bibfnamefont {T.~M.}\ \bibnamefont
  {Hoang}}, \bibinfo {author} {\bibfnamefont {J.}~\bibnamefont {Ahn}}, \bibinfo
  {author} {\bibfnamefont {J.}~\bibnamefont {Bang}}, \ and\ \bibinfo {author}
  {\bibfnamefont {T.}~\bibnamefont {Li}},\ }\href@noop {} {\bibfield  {journal}
  {\bibinfo  {journal} {Nature communications}\ }\textbf {\bibinfo {volume}
  {7}},\ \bibinfo {pages} {12250} (\bibinfo {year} {2016})}\BibitemShut
  {NoStop}%
\end{thebibliography}%

%
%
%

\end{document}